\documentclass[conference]{IEEEtran}

\usepackage{amsmath,amssymb,amsfonts}
\usepackage{graphicx}
\usepackage{cite}
\usepackage{booktabs}
\usepackage{array}
\usepackage{multirow}

\usepackage{url}
\usepackage{xcolor}
\usepackage{float}
\usepackage{stfloats}
\usepackage{booktabs}
\usepackage{array}
\usepackage{subcaption}

\DeclareMathOperator{\PV}{P.V.}

\begin{document}

\title{Two-Dimensional Method-of-Moments Analysis of TM$_z$ and TE$_z$ Scattering from PEC Cylinders}

\author{
\IEEEauthorblockN{Sabrina Saima}
\IEEEauthorblockA{
Elmore Family School of Electrical and Computer Engineering\\
Purdue University\\
West Lafayette, IN, USA\\
ssaima@purdue.edu}
}

\maketitle

\begin{abstract}
This paper presents a two-dimensional method-of-moments (MoM) solver for electromagnetic scattering from infinitely long perfectly electrically conducting (PEC) cylinders. Both TM$_z$ and TE$_z$ polarizations are considered. Starting from the scalar Helmholtz equation, the electric field integral equation (EFIE) is derived for TM$_z$ scattering and the magnetic field integral equation (MFIE) is derived for TE$_z$ scattering. The induced surface current on the PEC boundary is expanded using pulse basis functions, and the boundary integral equations are discretized using point matching at the segment centers. Circular cylinders with radii $R=\lambda$ and $R=2\lambda$ are used as validation cases because analytical series solutions are available. The MoM-computed surface currents, total near fields, scattered near fields, and field-error distributions are compared against the analytical solutions. After validation, the same solver is applied to a square PEC cylinder, for which no simple closed-form analytical solution is used. The results show strong agreement between the MoM and analytical circular-cylinder solutions and demonstrate the geometry-dependent scattering behavior of the square cylinder.
\end{abstract}


\section{Introduction}

\IEEEPARstart{E}{lectromagnetic} scattering from perfectly electrically conducting (PEC) objects is a classical problem in computational electromagnetics and is important in applications such as radar cross-section prediction and wave-object interaction analysis. Although analytical solutions exist for only a limited set of canonical geometries, most practical scattering problems require numerical methods. For open-region scattering problems, the method of moments (MoM) is especially attractive because it reformulates Maxwell's equations as boundary integral equations, so that only the scatterer boundary must be discretized and the radiation condition is automatically satisfied through the Green's function formulation \cite{harrington1993,gibson2008,jin2011}.

Other numerical techniques, such as the finite-difference time-domain (FDTD) method and the finite element method (FEM), are also widely used in electromagnetics \cite{jin_fem,taflove2005,saima2026waveguide}. In applied electromagnetics and photonics, these tools are routinely used for device-level modeling, including FDTD-based resonator sensing and FEM-based waveguide/optical simulations \cite{intisar2024soi,saima2023mim,zahin2026pra}. However, for open-region scattering, FDTD and FEM generally require volumetric discretization of the surrounding space and an artificial absorbing boundary condition or perfectly matched layer to emulate free space \cite{harrington1993,gibson2008,jin2011}. In contrast, MoM is naturally suited to this problem because the radiation condition is built into the free-space Green's function, making it an efficient approach for scattering from PEC cylinders.

In this work, the MoM formulation is applied to scattering from infinitely long PEC cylinders under plane-wave illumination, reducing the problem to a two-dimensional boundary-value problem in TM$_z$ and TE$_z$ polarization. For TM$_z$, the nonzero field component is $E_z$, and the PEC boundary condition leads to an electric field integral equation (EFIE). For TE$_z$, the nonzero field component is $H_z$, and the PEC boundary condition leads to a magnetic field integral equation (MFIE). These equations are discretized using pulse basis functions and point matching on a segmented boundary, producing dense linear systems whose solutions give the induced surface currents \cite{gibson2008,jin2011}.

The solver is first validated using circular PEC cylinders, since analytical cylindrical-wave series solutions are available for both polarizations. After validation, the same code is applied to a square PEC cylinder to demonstrate the ability of the MoM formulation to handle non-canonical geometries and to study the effects of edges and corners on the scattered fields \cite{harrington1993,gibson2008}.

The remainder of this paper is organized as follows. Section II presents the formulation and method-of-moments discretization, beginning with the governing equations, two-dimensional reduction, Green's function, and boundary integral representation. The section then derives the TM$_z$ EFIE and TE$_z$ MFIE, followed by the pulse-basis point-matching discretization and the analytical circular-cylinder solution used for validation. Section III describes the simulation setup and numerical parameters. Section IV presents the results for the circular-cylinder validation cases with $R=\lambda$ and $R=2\lambda$, as well as the scattering results for the square cylinder. Finally, Section V concludes the paper and discusses possible improvements and extensions.
\section{Formulation and Method-of-Moments Discretization}

\subsection{Governing Equations and Two-Dimensional Reduction}

We consider time-harmonic electromagnetic fields with an $e^{j\omega t}$
time convention. In the homogeneous free-space region exterior to the
scatterer, Maxwell's curl equations are
\begin{align}
\nabla \times \mathbf{E} &= -j\omega\mu_0 \mathbf{H},
\label{eq:faraday_mom}\\
\nabla \times \mathbf{H} &= j\omega\varepsilon_0 \mathbf{E},
\label{eq:ampere_mom}
\end{align}
where $\varepsilon_0$ and $\mu_0$ are the free-space permittivity and
permeability. Taking the curl of \eqref{eq:faraday_mom} and using
\eqref{eq:ampere_mom} gives the vector Helmholtz equation
\begin{equation}
\nabla^2 \mathbf{E}+k_0^2\mathbf{E}=\mathbf{0},
\label{eq:vector_helmholtz_E}
\end{equation}
where
\begin{equation}
k_0=\omega\sqrt{\mu_0\varepsilon_0}=\frac{2\pi}{\lambda}
\label{eq:k0_def_mom}
\end{equation}
is the free-space wavenumber. An identical vector Helmholtz equation is
obtained for $\mathbf{H}$.

The objects considered in this work are infinitely long PEC cylinders
whose axes are parallel to $\hat{z}$. Therefore, the geometry and fields
are invariant in the $z$ direction:
\begin{equation}
\frac{\partial}{\partial z}(\cdot)=0.
\label{eq:z_invariance_mom}
\end{equation}
Under this assumption, the electromagnetic field separates into two
independent polarizations. For TM$_z$ polarization, the relevant scalar
unknown is $E_z$, while for TE$_z$ polarization, the relevant scalar
unknown is $H_z$. Thus, we define
\begin{equation}
\psi(\boldsymbol{\rho}) =
\begin{cases}
E_z(\boldsymbol{\rho}), & \text{TM}_z,\\
H_z(\boldsymbol{\rho}), & \text{TE}_z,
\end{cases}
\label{eq:psi_definition}
\end{equation}
where
\begin{equation}
\boldsymbol{\rho}=x\hat{\mathbf{x}}+y\hat{\mathbf{y}}
\label{eq:rho_definition_mom}
\end{equation}
is the two-dimensional position vector. In the exterior region
$\Omega_\infty$ outside the PEC cylinder, both scalar fields satisfy the
two-dimensional Helmholtz equation
\begin{equation}
\nabla_t^2\psi(\boldsymbol{\rho})
+k_0^2\psi(\boldsymbol{\rho})=0,
\qquad
\boldsymbol{\rho}\in\Omega_\infty,
\label{eq:scalar_helmholtz_mom}
\end{equation}
where $\nabla_t^2$ is the transverse Laplacian in the $x$--$y$ plane.

The total field is written as the sum of incident and scattered fields,
\begin{equation}
\psi^{\mathrm{tot}}(\boldsymbol{\rho})
=
\psi^{\mathrm{inc}}(\boldsymbol{\rho})
+
\psi^{\mathrm{scat}}(\boldsymbol{\rho}).
\label{eq:total_field_decomposition_mom}
\end{equation}
The scattered field must satisfy the two-dimensional Sommerfeld radiation
condition,
\begin{equation}
\lim_{\rho\rightarrow\infty}
\sqrt{\rho}
\left(
\frac{\partial \psi^{\mathrm{scat}}}{\partial \rho}
+jk_0\psi^{\mathrm{scat}}
\right)=0,
\label{eq:sommerfeld_mom}
\end{equation}
which enforces outward-radiating cylindrical waves at infinity.

\subsection{Two-Dimensional Green's Function}

The two-dimensional free-space Green's function used in this work satisfies
\begin{equation}
\nabla_t^2 G_0(\boldsymbol{\rho},\boldsymbol{\rho}')
+k_0^2G_0(\boldsymbol{\rho},\boldsymbol{\rho}')
=
-\delta(\boldsymbol{\rho}-\boldsymbol{\rho}'),
\label{eq:green_eq_mom}
\end{equation}
together with the outgoing-wave radiation condition. The corresponding
solution is
\begin{equation}
G_0(\boldsymbol{\rho},\boldsymbol{\rho}')
=
-\frac{j}{4}
H_0^{(2)}
\left(k_0|\boldsymbol{\rho}-\boldsymbol{\rho}'|\right),
\label{eq:green_func_mom}
\end{equation}
where $H_0^{(2)}(\cdot)$ is the zeroth-order Hankel function of the second
kind. This Green's function represents the field radiated by a two-dimensional
point source and is the kernel that couples every source point on the PEC
boundary to every observation point.

For the TE$_z$ formulation, the normal derivative of the Green's function
is also required. Applying the chain rule and using
\begin{equation}
\frac{d}{dz}H_0^{(2)}(z)=-H_1^{(2)}(z),
\label{eq:hankel_derivative_identity}
\end{equation}
gives
\begin{equation}
\frac{\partial G_0(\boldsymbol{\rho},\boldsymbol{\rho}')}{\partial n'}
=
\frac{k_0}{4j}
H_1^{(2)}
\left(k_0|\boldsymbol{\rho}-\boldsymbol{\rho}'|\right)
\frac{
\hat{\mathbf{n}}'\cdot
(\boldsymbol{\rho}-\boldsymbol{\rho}')
}{
|\boldsymbol{\rho}-\boldsymbol{\rho}'|
}.
\label{eq:dGdn_mom}
\end{equation}
Here $\hat{\mathbf{n}}'$ is the outward unit normal at the source point
$\boldsymbol{\rho}'$ on the PEC boundary.

\subsection{Boundary Integral Representation}

Let $\Gamma_o$ denote the boundary contour of the PEC cylinder. Applying
Green's second identity to $\psi$ and $G_0$ in the exterior region and then
bringing the observation point to the boundary gives the surface integral
representation for $\boldsymbol{\rho}\in\Gamma_o^{-}$ is
\begin{equation}
\begin{aligned}
\psi^{\mathrm{inc}}(\boldsymbol{\rho})
&+
\int_{\Gamma_o}
\bigg[
\psi(\boldsymbol{\rho}')
\frac{\partial G_0(\boldsymbol{\rho},\boldsymbol{\rho}')}{\partial n'}
\\
&\qquad
-
G_0(\boldsymbol{\rho},\boldsymbol{\rho}')
\frac{\partial \psi(\boldsymbol{\rho}')}{\partial n'}
\bigg]\, d\ell'
=
\frac{1}{2}\psi(\boldsymbol{\rho}).
\end{aligned}
\label{eq:general_bie_mom}
\end{equation}

The notation $\Gamma_o^{-}$ indicates that the observation point is taken
just inside the boundary during the limiting process. The factor
$\frac{1}{2}$ appears because the singular part of the integral operator
must be evaluated in the principal-value sense as the observation point is
brought onto a smooth boundary. Equation \eqref{eq:general_bie_mom} is the
starting point for deriving the specific integral equations used for the
TM$_z$ and TE$_z$ cases.

\subsection{TM$_z$ Electric Field Integral Equation}

For TM$_z$ polarization,
\begin{equation}
\mathbf{E}(\boldsymbol{\rho})
=
\hat{\mathbf{z}}E_z(\boldsymbol{\rho}),
\label{eq:tmz_E_field}
\end{equation}
and the magnetic field lies in the transverse plane. Since $E_z$ is
tangential to the surface of an infinitely long PEC cylinder, the PEC
boundary condition requires
\begin{equation}
E_z^{\mathrm{tot}}(\boldsymbol{\rho})=0,
\qquad
\boldsymbol{\rho}\in\Gamma_o.
\label{eq:tmz_pec_bc}
\end{equation}
Equivalently,
\begin{equation}
E_z^{\mathrm{inc}}(\boldsymbol{\rho})
+
E_z^{\mathrm{scat}}(\boldsymbol{\rho})
=0,
\qquad
\boldsymbol{\rho}\in\Gamma_o.
\label{eq:tmz_total_bc}
\end{equation}

For the TM$_z$ problem, the induced surface current is $z$ directed and is
denoted by $J_{s,z}$. The boundary relation between the normal derivative
of $E_z$ and the induced current is
\begin{equation}
\frac{\partial E_z}{\partial n'}
=
jk_0Z_0J_{s,z},
\qquad
\boldsymbol{\rho}'\in\Gamma_o,
\label{eq:tmz_neumann_current}
\end{equation}
where
\begin{equation}
Z_0=\sqrt{\frac{\mu_0}{\varepsilon_0}}=120\pi~\Omega
\label{eq:z0_def_mom}
\end{equation}
is the free-space wave impedance. Substituting the PEC condition
$E_z=0$ and \eqref{eq:tmz_neumann_current} into
\eqref{eq:general_bie_mom} yields the TM$_z$ electric field integral
equation (EFIE):
\begin{equation}
E_z^{\mathrm{inc}}(\boldsymbol{\rho})
-
jk_0Z_0
\int_{\Gamma_o}
G_0(\boldsymbol{\rho},\boldsymbol{\rho}')
J_{s,z}(\boldsymbol{\rho}')
d\Gamma'
=
0,
\qquad
\boldsymbol{\rho}\in\Gamma_o.
\label{eq:tmz_efie_mom}
\end{equation}
This equation has a direct physical interpretation: the incident electric
field on the PEC surface is exactly cancelled by the scattered electric
field radiated by the induced surface current. Therefore, the total
tangential electric field on the PEC boundary is forced to zero.

\subsection{TE$_z$ Magnetic Field Integral Equation}

For TE$_z$ polarization,
\begin{equation}
\mathbf{H}(\boldsymbol{\rho})
=
\hat{\mathbf{z}}H_z(\boldsymbol{\rho}),
\label{eq:tez_H_field}
\end{equation}
and the electric field lies in the transverse plane. The PEC boundary
condition requires the tangential electric field to vanish. In the scalar
TE$_z$ formulation, this gives the homogeneous Neumann condition
\begin{equation}
\frac{\partial H_z^{\mathrm{tot}}}{\partial n}=0,
\qquad
\boldsymbol{\rho}\in\Gamma_o.
\label{eq:tez_neumann_bc}
\end{equation}
The induced surface current is tangential to the boundary and is related
to the magnetic field at the PEC surface by
\begin{equation}
H_z(\boldsymbol{\rho})=-J_{s,t}(\boldsymbol{\rho}),
\qquad
\boldsymbol{\rho}\in\Gamma_o,
\label{eq:tez_current_relation}
\end{equation}
where $J_{s,t}$ is the tangential surface current density.

Substituting \eqref{eq:tez_neumann_bc} and
\eqref{eq:tez_current_relation} into the boundary integral representation
\eqref{eq:general_bie_mom} gives the TE$_z$ magnetic field integral
equation (MFIE):
\begin{equation}
H_z^{\mathrm{inc}}(\boldsymbol{\rho})
-
\PV\int_{\Gamma_o}
\frac{\partial G_0(\boldsymbol{\rho},\boldsymbol{\rho}')}{\partial n'}
J_{s,t}(\boldsymbol{\rho}')
d\Gamma'
=
-\frac{1}{2}J_{s,t}(\boldsymbol{\rho}).
\label{eq:tez_mfie_mom}
\end{equation}
Equivalently, this may be written as
\begin{equation}
-\frac{1}{2}J_{s,t}(\boldsymbol{\rho})
+
\PV\int_{\Gamma_o}
\frac{\partial G_0(\boldsymbol{\rho},\boldsymbol{\rho}')}{\partial n'}
J_{s,t}(\boldsymbol{\rho}')
d\Gamma'
=
H_z^{\mathrm{inc}}(\boldsymbol{\rho}).
\label{eq:tez_mfie_matrix_form}
\end{equation}
Unlike the EFIE, the MFIE is an integral equation of the second kind
because the unknown current appears both inside and outside the integral
operator. This distinction is important numerically: second-kind integral
equations often have different conditioning behavior than first-kind
integral equations, although accurate evaluation of the principal-value
operator is still required.

\subsection{Boundary Segmentation and Pulse Basis Functions}

The continuous boundary $\Gamma_o$ is divided into $N$ straight-line
segments,
\begin{equation}
\Gamma_o=\bigcup_{n=1}^{N}s_n.
\label{eq:boundary_segmentation_final}
\end{equation}
The center of the $n$th segment is denoted by $\boldsymbol{\rho}_n$, and
its length is denoted by $\Delta \ell_n$. For a circular cylinder, the
segments are uniformly distributed in the angular coordinate. For the
square cylinder, each side is divided into straight segments and the full
contour is traversed counterclockwise.

The unknown surface current is approximated as constant over each
segment using pulse basis functions. The $n$th pulse basis function is
defined as
\begin{equation}
p_n(\boldsymbol{\rho})=
\begin{cases}
1, & \boldsymbol{\rho}\in s_n,\\
0, & \text{otherwise}.
\end{cases}
\label{eq:pulse_basis_final}
\end{equation}
Thus, for TM$_z$ polarization,
\begin{equation}
J_{s,z}(\boldsymbol{\rho})
\approx
\sum_{n=1}^{N}I_n^{\mathrm{TM}}p_n(\boldsymbol{\rho}),
\label{eq:tmz_pulse_expansion_final}
\end{equation}
and for TE$_z$ polarization,
\begin{equation}
J_{s,t}(\boldsymbol{\rho})
\approx
\sum_{n=1}^{N}I_n^{\mathrm{TE}}p_n(\boldsymbol{\rho}).
\label{eq:tez_pulse_expansion_final}
\end{equation}
The integral equations are enforced at the segment centers,
\begin{equation}
\boldsymbol{\rho}=\boldsymbol{\rho}_m,
\qquad
m=1,2,\ldots,N.
\label{eq:point_matching_final}
\end{equation}
This testing procedure is called point matching or collocation. It
converts the continuous boundary integral equations into dense linear
systems of the form
\begin{equation}
\sum_{n=1}^{N}Z_{mn}I_n=V_m,
\qquad
m=1,2,\ldots,N,
\label{eq:mom_linear_system_final}
\end{equation}
or, in matrix form,
\begin{equation}
[\mathbf{Z}]\{\mathbf{I}\}=\{\mathbf{V}\}.
\label{eq:mom_matrix_form_final}
\end{equation}
Because the Green's function is nonlocal, every segment interacts with
every other segment, so the MoM impedance matrix is dense.

\subsection{TM$_z$ MoM Matrix}

Substituting the pulse expansion \eqref{eq:tmz_pulse_expansion_final}
into the EFIE \eqref{eq:tmz_efie_mom} and applying point matching gives
\begin{equation}
\sum_{n=1}^{N}
Z_{mn}^{\mathrm{TM}}I_n^{\mathrm{TM}}
=
V_m^{\mathrm{TM}},
\qquad
m=1,2,\ldots,N.
\label{eq:tmz_matrix_system_final}
\end{equation}
The excitation vector is
\begin{equation}
V_m^{\mathrm{TM}}
=
E_z^{\mathrm{inc}}(\boldsymbol{\rho}_m).
\label{eq:tmz_rhs_final}
\end{equation}
For a unit-amplitude plane wave propagating in the $+x$ direction,
\begin{equation}
E_z^{\mathrm{inc}}(\boldsymbol{\rho}_m)
=
e^{-jk_0x_m}.
\label{eq:tmz_plane_wave_rhs_final}
\end{equation}

The matrix entries are
\begin{equation}
Z_{mn}^{\mathrm{TM}}
=
jk_0Z_0
\int_{s_n}
G_0(\boldsymbol{\rho}_m,\boldsymbol{\rho}')
d\Gamma'.
\label{eq:tmz_zmn_integral_final}
\end{equation}
For $m\neq n$, the integrand is nonsingular and is evaluated using the
midpoint approximation:
\begin{equation}
Z_{mn}^{\mathrm{TM}}
\approx
\frac{k_0Z_0\Delta \ell_n}{4}
H_0^{(2)}
\left(k_0|\boldsymbol{\rho}_m-\boldsymbol{\rho}_n|\right),
\qquad
m\neq n.
\label{eq:tmz_offdiag_final}
\end{equation}
For $m=n$, the Green's function is singular. Using the small-argument
expansion of the Hankel function and integrating over the segment gives
\begin{equation}
Z_{mm}^{\mathrm{TM}}
\approx
\frac{k_0Z_0\Delta \ell_m}{4}
\left[
1
-
j\frac{2}{\pi}
\ln
\left(
\frac{k_0\gamma\Delta \ell_m}{4e}
\right)
\right],
\label{eq:tmz_self_final}
\end{equation}
where $\gamma=e^{\gamma_E}\approx1.781$, $\gamma_E$ is the
Euler--Mascheroni constant, and $e\approx2.7183$ is Euler's number.
This special diagonal treatment is necessary because the midpoint rule
cannot be applied directly at the singularity.

\subsection{TE$_z$ MoM Matrix}

Similarly, substituting the pulse expansion
\eqref{eq:tez_pulse_expansion_final} into the MFIE
\eqref{eq:tez_mfie_matrix_form} and applying point matching gives
\begin{equation}
\sum_{n=1}^{N}
Z_{mn}^{\mathrm{TE}}I_n^{\mathrm{TE}}
=
V_m^{\mathrm{TE}},
\qquad
m=1,2,\ldots,N.
\label{eq:tez_matrix_system_final}
\end{equation}
The right-hand side is
\begin{equation}
V_m^{\mathrm{TE}}
=
H_z^{\mathrm{inc}}(\boldsymbol{\rho}_m).
\label{eq:tez_rhs_final}
\end{equation}
For a unit-amplitude plane wave traveling in the $+x$ direction,
\begin{equation}
H_z^{\mathrm{inc}}(\boldsymbol{\rho}_m)
=
e^{-jk_0x_m}.
\label{eq:tez_plane_wave_rhs_final}
\end{equation}

For $m\neq n$, the matrix entries are obtained by midpoint approximation
of the normal-derivative kernel:
\begin{equation}
Z_{mn}^{\mathrm{TE}}
\approx
\frac{k_0\Delta \ell_n}{4j}
H_1^{(2)}
\left(k_0|\boldsymbol{\rho}_m-\boldsymbol{\rho}_n|\right)
\frac{
\hat{\mathbf{n}}_n\cdot
(\boldsymbol{\rho}_m-\boldsymbol{\rho}_n)
}{
|\boldsymbol{\rho}_m-\boldsymbol{\rho}_n|
}.
\label{eq:tez_offdiag_final}
\end{equation}
For the diagonal term, the principal-value contribution of the
normal-derivative integral vanishes for a smooth or locally flat segment.
Therefore, only the identity contribution from the MFIE remains:
\begin{equation}
Z_{mm}^{\mathrm{TE}}=-\frac{1}{2}.
\label{eq:tez_self_final}
\end{equation}

\subsection{Near-Field Computation}

After solving \eqref{eq:tmz_matrix_system_final} or
\eqref{eq:tez_matrix_system_final}, the resulting surface current is used
to reconstruct the scattered and total fields in the exterior region.

For TM$_z$ polarization, the scattered electric field is
\begin{equation}
E_z^{\mathrm{scat}}(\boldsymbol{\rho})
=
-jk_0Z_0
\int_{\Gamma_o}
G_0(\boldsymbol{\rho},\boldsymbol{\rho}')
J_{s,z}(\boldsymbol{\rho}')
d\Gamma'.
\label{eq:tmz_scattered_integral_final}
\end{equation}
Using the pulse expansion and midpoint approximation gives
\begin{equation}
E_z^{\mathrm{scat}}(\boldsymbol{\rho})
\approx
-\frac{k_0Z_0}{4}
\sum_{n=1}^{N}
I_n^{\mathrm{TM}}\Delta \ell_n
H_0^{(2)}
\left(k_0|\boldsymbol{\rho}-\boldsymbol{\rho}_n|\right).
\label{eq:tmz_scattered_discrete_final}
\end{equation}
The total field is then
\begin{equation}
E_z^{\mathrm{tot}}(\boldsymbol{\rho})
=
E_z^{\mathrm{inc}}(\boldsymbol{\rho})
+
E_z^{\mathrm{scat}}(\boldsymbol{\rho}).
\label{eq:tmz_total_final}
\end{equation}

For TE$_z$ polarization, the scattered magnetic field is reconstructed as
\begin{equation}
H_z^{\mathrm{scat}}(\boldsymbol{\rho})
=
-\int_{\Gamma_o}
\frac{\partial G_0(\boldsymbol{\rho},\boldsymbol{\rho}')}{\partial n'}
J_{s,t}(\boldsymbol{\rho}')
d\Gamma',
\label{eq:tez_scattered_integral_final}
\end{equation}
which becomes
\begin{equation}
H_z^{\mathrm{scat}}(\boldsymbol{\rho})
\approx
-
\sum_{n=1}^{N}
I_n^{\mathrm{TE}}\Delta \ell_n
\frac{\partial G_0(\boldsymbol{\rho},\boldsymbol{\rho}_n)}
{\partial n'}.
\label{eq:tez_scattered_discrete_final}
\end{equation}
The total magnetic field is
\begin{equation}
H_z^{\mathrm{tot}}(\boldsymbol{\rho})
=
H_z^{\mathrm{inc}}(\boldsymbol{\rho})
+
H_z^{\mathrm{scat}}(\boldsymbol{\rho}).
\label{eq:tez_total_final}
\end{equation}
The negative signs in \eqref{eq:tmz_scattered_integral_final} and
\eqref{eq:tez_scattered_integral_final} are important and are consistent
with the scattered-field evaluation used in the MATLAB implementation.

\subsection{Analytical Circular-Cylinder Validation}

For a circular PEC cylinder of radius $a$, the MoM solution can be
validated using the analytical cylindrical-wave series solution. A plane
wave traveling in the $+x$ direction can be expanded as
\begin{equation}
\psi^{\mathrm{inc}}(\rho,\phi)
=
e^{-jk_0x}
=
\sum_{n=-\infty}^{\infty}
(-j)^nJ_n(k_0\rho)e^{jn\phi},
\label{eq:incident_mie_series_final}
\end{equation}
where $J_n(\cdot)$ is the Bessel function of the first kind. The scattered
field is written as
\begin{equation}
\psi^{\mathrm{scat}}(\rho,\phi)
=
\sum_{n=-\infty}^{\infty}
c_nH_n^{(2)}(k_0\rho)e^{jn\phi},
\label{eq:scattered_mie_series_final}
\end{equation}
where the coefficients $c_n$ are determined by the PEC boundary condition.

For TM$_z$ polarization, the boundary condition is
$E_z^{\mathrm{tot}}=0$ at $\rho=a$, giving
\begin{equation}
c_n^{\mathrm{TM}}
=
-(-j)^n
\frac{J_n(k_0a)}{H_n^{(2)}(k_0a)}.
\label{eq:tmz_mie_coeff_final}
\end{equation}
For TE$_z$ polarization, the boundary condition is
$\partial H_z^{\mathrm{tot}}/\partial \rho=0$ at $\rho=a$, giving
\begin{equation}
c_n^{\mathrm{TE}}
=
-(-j)^n
\frac{J_n'(k_0a)}{H_n^{(2)\prime}(k_0a)}.
\label{eq:tez_mie_coeff_final}
\end{equation}
The derivative is evaluated using
\begin{equation}
Z_n'(x)=\frac{1}{2}\left[Z_{n-1}(x)-Z_{n+1}(x)\right],
\label{eq:bessel_derivative_recurrence_final}
\end{equation}
where $Z_n$ may represent either $J_n$ or $H_n^{(2)}$.

The analytical TM$_z$ surface current on the circular PEC cylinder is
\begin{equation}
J_{s,z}^{\mathrm{ana}}(\phi)
=
\frac{2}{\pi k_0Z_0a}
\sum_{n=-\infty}^{\infty}
\frac{j^{-n}}{H_n^{(2)}(k_0a)}
e^{jn\phi}.
\label{eq:tmz_analytical_current_final}
\end{equation}
For TE$_z$ polarization, the analytical surface current is obtained from
\begin{equation}
J_{s,t}^{\mathrm{ana}}(\phi)
=
-H_z^{\mathrm{tot}}(a,\phi).
\label{eq:tez_analytical_current_final}
\end{equation}
In the numerical implementation, the infinite series is truncated to
$|n|\leq M_{\max}$. The truncation parameter is chosen large enough so that
the analytical current and near-field solutions are converged for the
electrical size of the cylinder.

\section{Simulation Setup}

All simulations presented in this work were implemented in
MATLAB\textsuperscript{\textregistered} (R2024a). The resulting dense linear systems were solved via the
built-in direct solver (backslash operator), which employs LU
factorization with partial pivoting.  No iterative solvers or
preconditioning strategies were necessary given the moderate problem sizes
considered. The background medium was free space and the excitation in every case is a unit-amplitude plane wave propagating in
the $+x$ direction:
\begin{equation}
  E_z^{\mathrm{inc}} = e^{-jk_0 x} \;\;(\text{TM}_z), \qquad
  H_z^{\mathrm{inc}} = e^{-jk_0 x} \;\;(\text{TE}_z).
\end{equation}
The scatterer boundary was discretized into straight segments and pulse
basis functions were used to approximate the induced surface current on
each segment. The integral equations were enforced at the segment centers
using point matching. Surface-current magnitudes were normalized by the
corresponding incident-field amplitude. In the plots, \(J_z/H_0\) is used
for the TM\(_z\) current and \(J_t/H_0\) is used for the TE\(_z\) current,
with \(H_0=1/Z_0\) for the TM\(_z\) normalization and \(H_0=1\) for the
TE\(_z\) normalization. Near-field quantities were normalized by the
magnitude of the corresponding incident scalar field. Observation points inside the
PEC objects were excluded from the field plots. For the square cylinder, a
small buffer region around the boundary was also excluded in the TE\(_z\)
field plots to suppress near-singular artifacts from the MFIE kernel. The main numerical parameters are summarized in Table~\ref{tab:simulation_parameters}.

\begin{table}[!t]
\caption{Simulation parameters for the MoM scattering study.}
\label{tab:simulation_parameters}
\centering
\footnotesize
\renewcommand{\arraystretch}{1.12}
\setlength{\tabcolsep}{5pt}
\begin{tabular}{@{}p{0.43\linewidth}p{0.47\linewidth}@{}}
\toprule
\multicolumn{2}{c}{\textbf{Circular PEC Cylinder: $R=\lambda$}} \\
\midrule
Radius & $R=\lambda$ \\
TM$_z$ segments & $N=200$ \\
TE$_z$ segments & $N=240$ \\
Series truncation & $M_{\max}=60$ (TM$_z$), $80$ (TE$_z$) \\
Plot window & $-5\lambda \leq x,y \leq 5\lambda$ \\
\midrule
\multicolumn{2}{c}{\textbf{Circular PEC Cylinder: $R=2\lambda$}} \\
\midrule
Radius & $R=2\lambda$ \\
Boundary segments & $N=350$ \\
Series truncation & $M_{\max}=80$ \\
Plot window & $-10\lambda \leq x,y \leq 10\lambda$ \\
\midrule
\multicolumn{2}{c}{\textbf{Square PEC Cylinder}} \\
\midrule
Side length & $3\lambda \times 3\lambda$ \\
Segments per side & $N_{\mathrm{side}}=30$ \\
Total segments & $N=120$ \\
Boundary parameter & $s/\lambda$, counterclockwise \\
Starting point & Midpoint of back/right face \\
Plot window & $-5\lambda \leq x,y \leq 5\lambda$ \\
\bottomrule
\end{tabular}
\end{table}


\section{Numerical Results}

The MoM implementation was first validated using circular PEC cylinders,
since closed-form cylindrical-wave solutions are available for both
TM$_z$ and TE$_z$ polarizations. Two cylinder radii were considered,
namely $R=\lambda$ and $R=2\lambda$. For each case, the numerical
surface current obtained from the MoM solution was first compared with
the corresponding analytical result. The near-field validation was then
performed by comparing the analytical and MoM total fields and scattered
fields.

\subsection{Circular PEC Cylinder with $R=\lambda$}

\subsubsection{TM$_z$ Polarization}

\begin{figure}[b]
\centering
\includegraphics[width=0.95\linewidth]{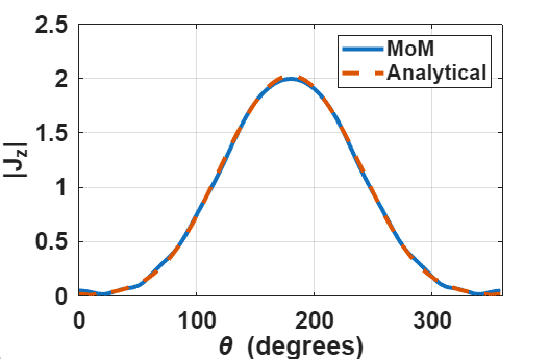}
\caption{Maginitude of the induced surface-current density for
TM$_z$ scattering from a circular PEC cylinder with $R=\lambda$.}
\label{fig:r1_tm_current1}
\end{figure}

We begin with TM$_z$ scattering from a circular PEC cylinder of radius
$R=\lambda$. Figure~\ref{fig:r1_tm_current1} compares the surface-current
magnitude obtained from the MoM solution with the analytical current
distribution. The agreement is very close over the full angular range,
which confirms that the EFIE formulation and its pulse-basis
discretization correctly capture the induced current on the PEC boundary.

\begin{figure}
\centerline{
\begin{tabular}{cc}
\includegraphics[width=0.49\linewidth]{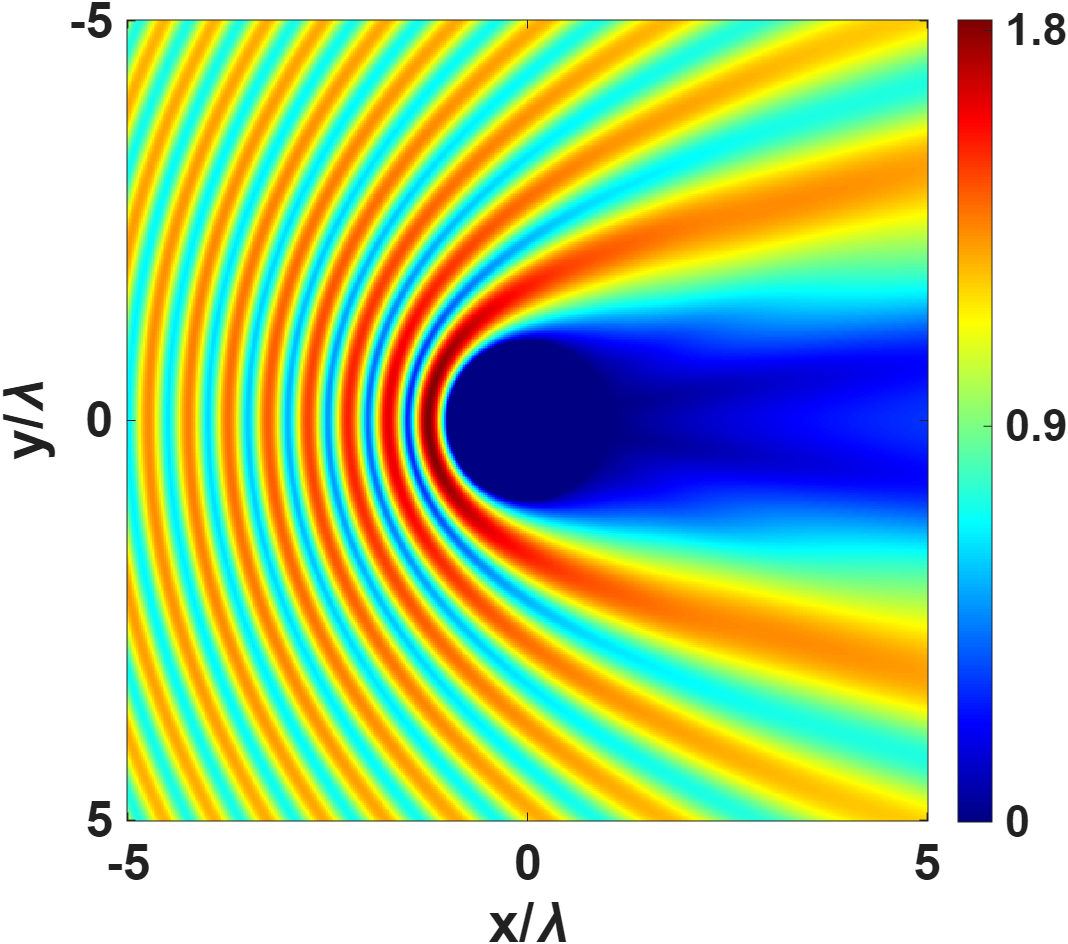} &
\includegraphics[width=0.49\linewidth]{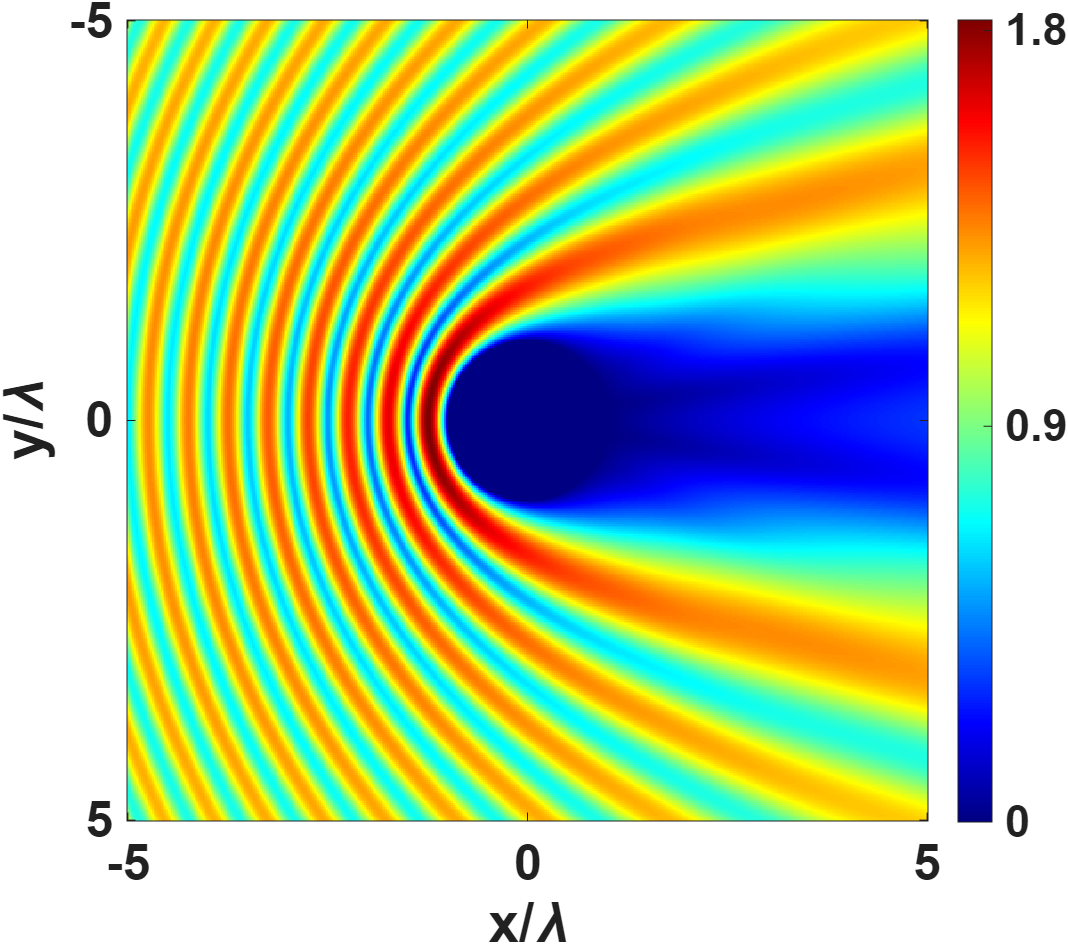} \\
{\centering\footnotesize (a) Analytical $|E_z^{\mathrm{tot}}|$} & {\centering\footnotesize (b) MoM $|E_z^{\mathrm{tot}}|$} \\
\end{tabular}
}
\vspace{2mm}
\centerline{
\begin{tabular}{cc}
\includegraphics[width=0.49\linewidth]{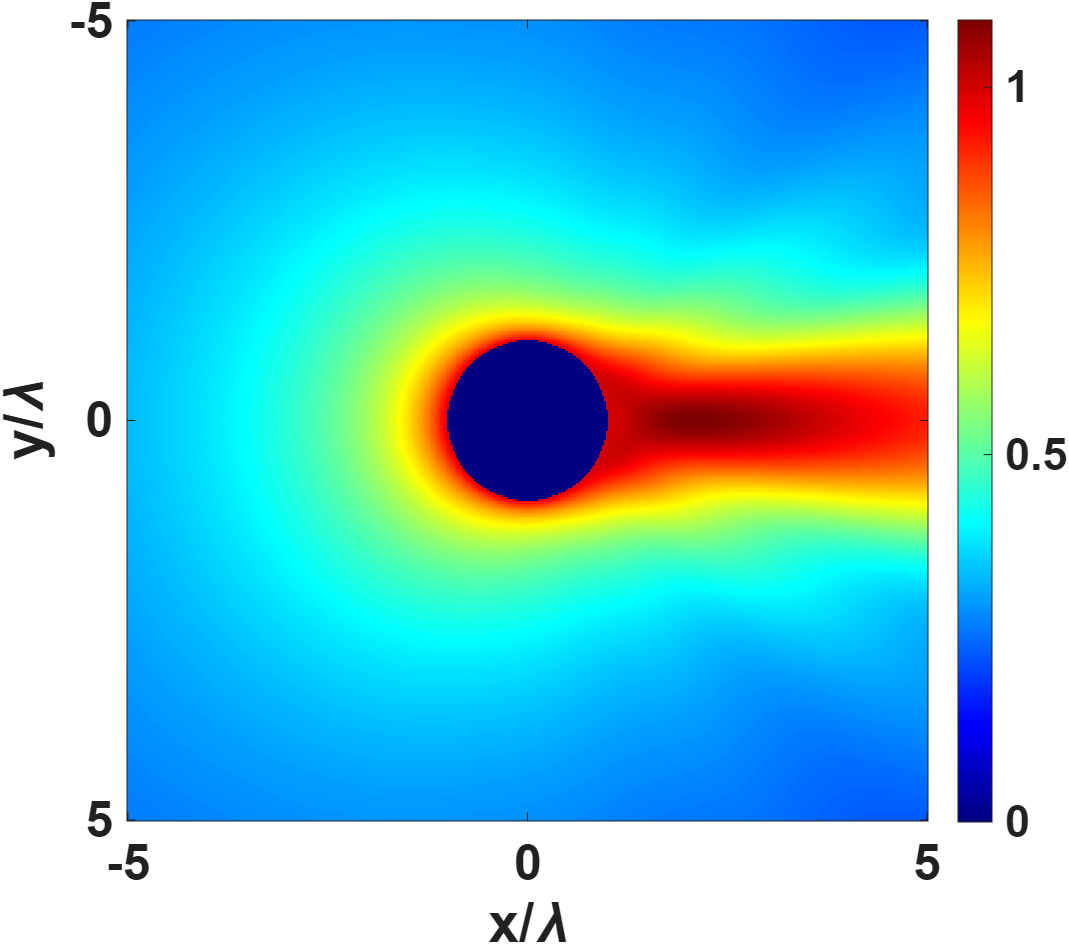} &
\includegraphics[width=0.49\linewidth]{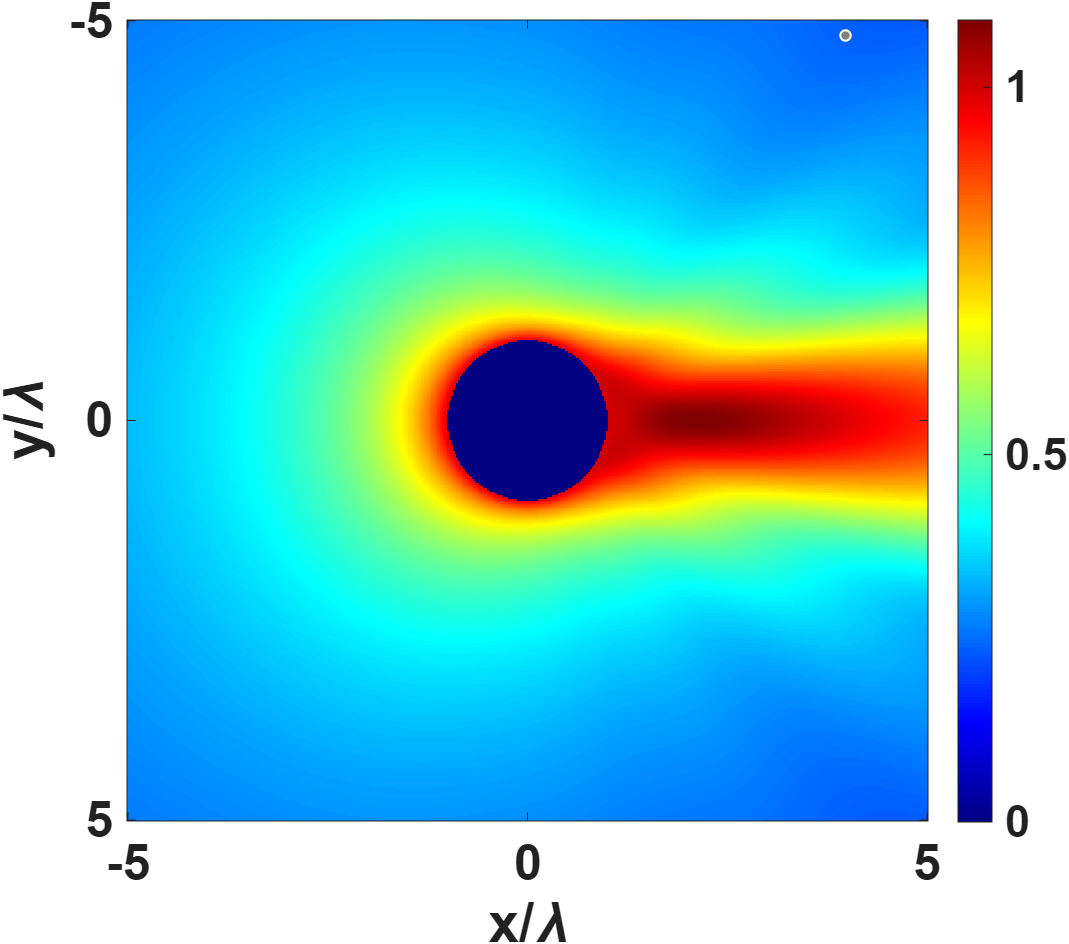} \\
{\centering\footnotesize (c) Analytical $|E_z^{\mathrm{scat}}|$} & {\centering\footnotesize (d) MoM $|E_z^{\mathrm{scat}}|$} \\
\end{tabular}
}
\caption{Analytical and MoM near-field comparison for TM$_z$ scattering
from a circular PEC cylinder with $R=\lambda$.}
\label{fig:r1_tm_fields}
\vspace{-5mm}
\end{figure}

Figure~\ref{fig:r1_tm_fields} compares the analytical and MoM near-field
solutions. The total-field plots in Fig.~\ref{fig:r1_tm_fields}(a) and
Fig.~\ref{fig:r1_tm_fields}(b) show the expected interference pattern in
front of the cylinder together with the shadow region behind the PEC
object. The scattered-field plots in
Fig.~\ref{fig:r1_tm_fields}(c)--(d) show that the MoM solution accurately
reproduces the field radiated by the induced current. The close visual
agreement between the analytical and numerical plots confirms the
correctness of the TM$_z$ implementation for the $R=\lambda$ case.

\subsubsection{TE$_z$ Polarization}

\begin{figure}[b]
\centering
\includegraphics[width=0.95\linewidth]{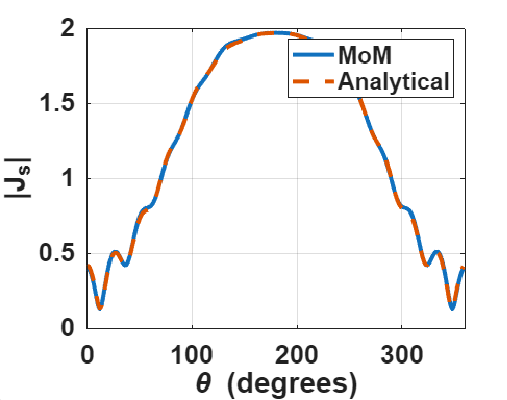}
\caption{Maginitude of the induced surface-current density for TE$_z$ scattering from a circular PEC cylinder with $R=\lambda$.}
\label{fig:r1_te_current}
\end{figure}

\begin{figure}
\centerline{
\begin{tabular}{cc}
\includegraphics[width=0.49\linewidth]{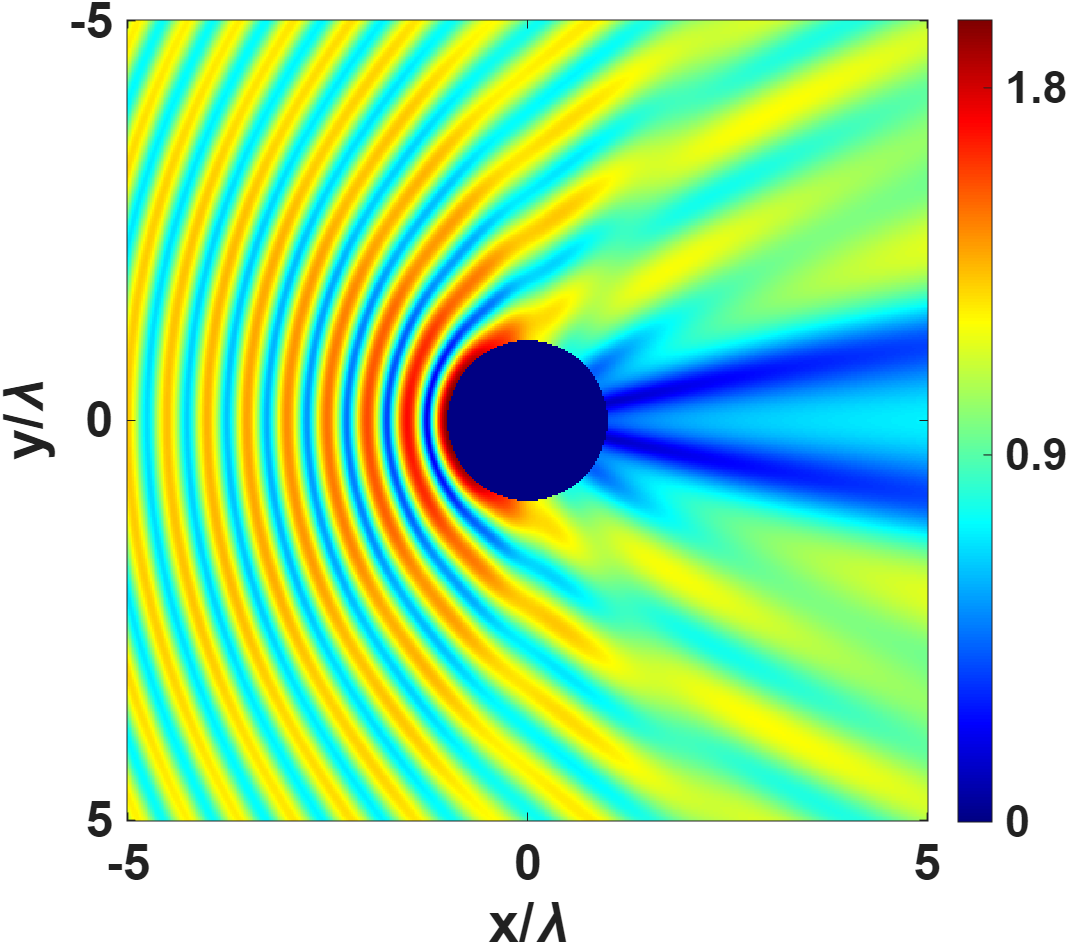} &
\includegraphics[width=0.49\linewidth]{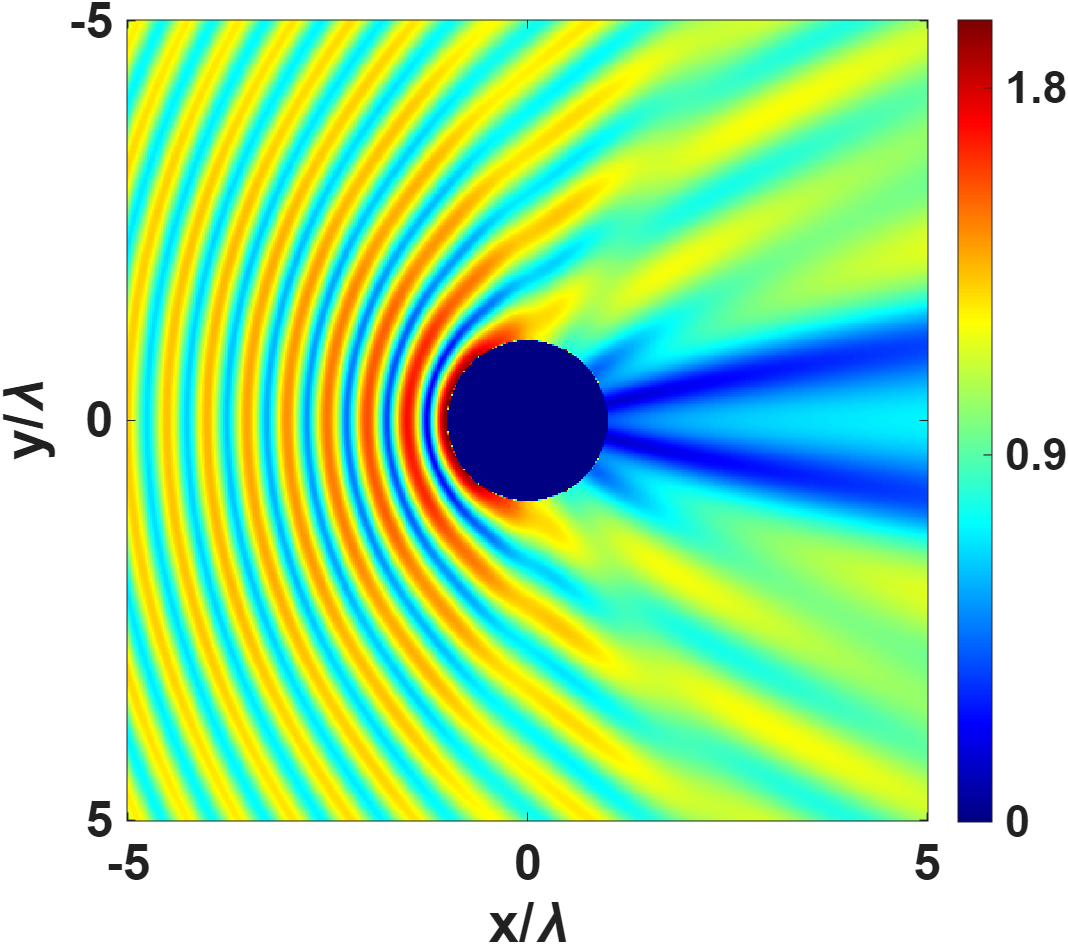} \\
{\centering\footnotesize (a) Analytical $|H_z^{\mathrm{tot}}|$} & {\centering\footnotesize (b) MoM $|H_z^{\mathrm{tot}}|$} \\
\end{tabular}
}
\vspace{2mm}
\centerline{
\begin{tabular}{cc}
\includegraphics[width=0.49\linewidth]{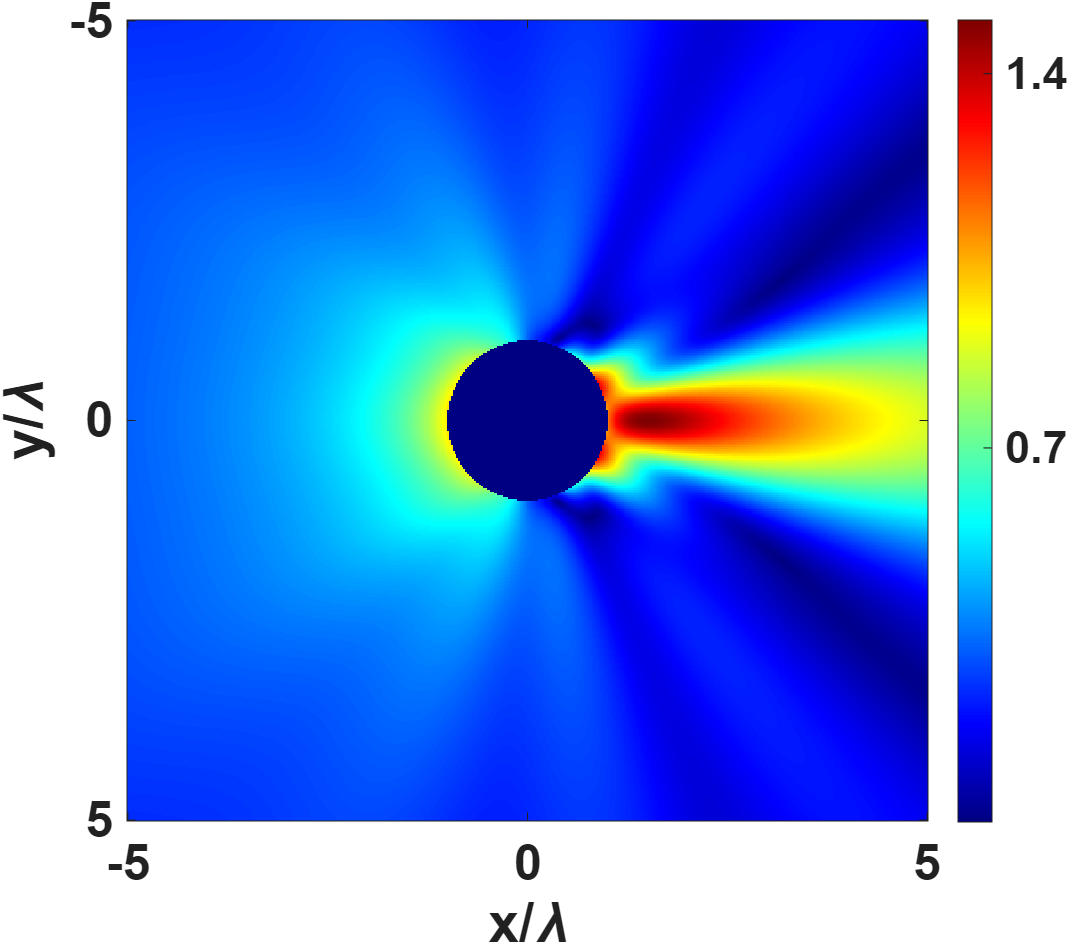} &
\includegraphics[width=0.49\linewidth]{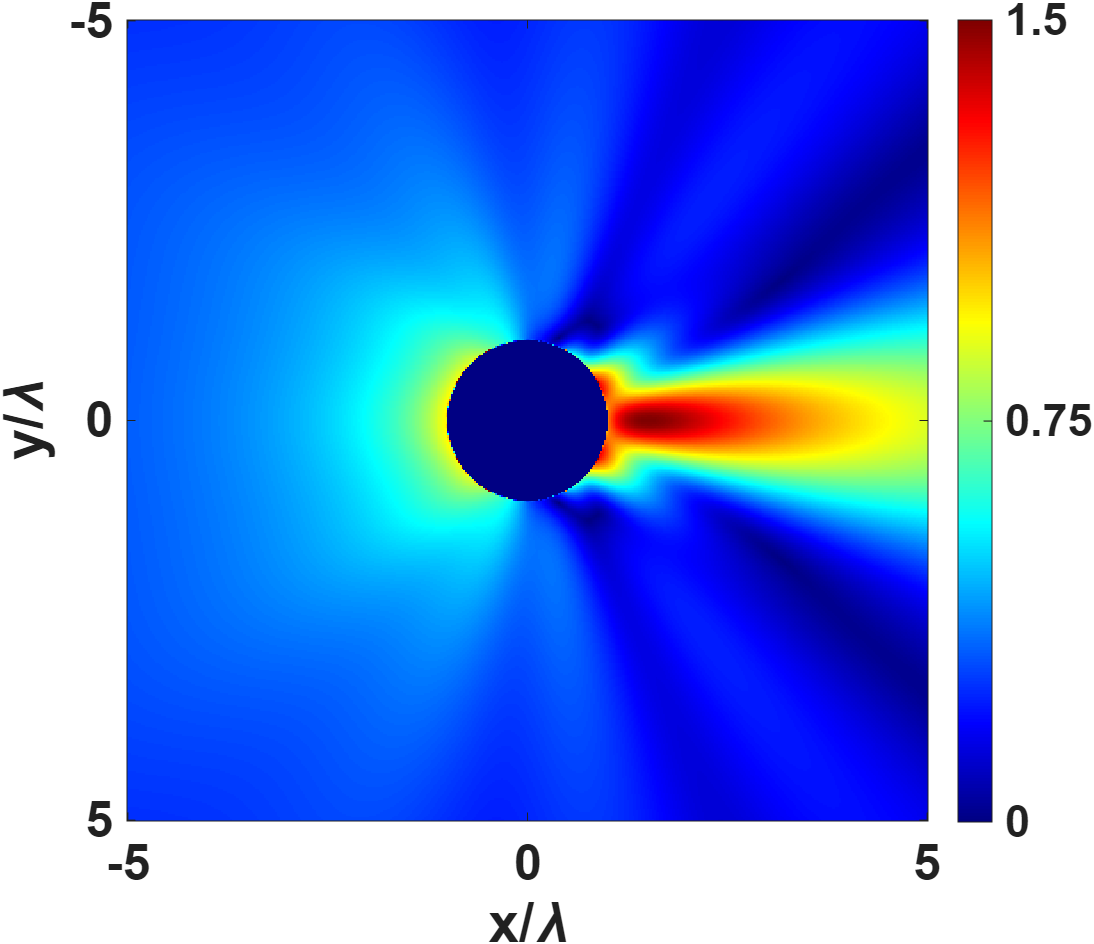} \\
{\centering\footnotesize (c) Analytical $|H_z^{\mathrm{scat}}|$} & {\centering\footnotesize (d) MoM $|H_z^{\mathrm{scat}}|$} \\
\end{tabular}
}
\caption{Analytical and MoM near-field comparison for TE$_z$ scattering
from a circular PEC cylinder with $R=\lambda$.}
\label{fig:r1_te_fields}
\vspace{-5mm}
\end{figure}

Next, the same circular PEC cylinder was analyzed for TE$_z$
polarization. Figure~\ref{fig:r1_te_current} compares the MoM and
analytical surface-current magnitudes. The two results agree well over
the full boundary, which confirms that the MFIE implementation correctly
captures the tangential surface current distribution.
The corresponding total-field and scattered-field comparisons are shown
in Fig.~\ref{fig:r1_te_fields}. The total magnetic field in
Fig.~\ref{fig:r1_te_fields}(a)--(b) again shows the expected illuminated
region and shadow region. The scattered-field plots in
Fig.~\ref{fig:r1_te_fields}(c)--(d) exhibit a stronger angular variation
than in the TM$_z$ case, which is consistent with the TE$_z$ boundary
condition and the normal-derivative kernel appearing in the MFIE. Overall,
the numerical and analytical field plots remain in very close agreement.

\subsection{Circular PEC Cylinder with $R=2\lambda$}

\subsubsection{TM$_z$ Polarization}

\begin{figure}[b!]
\centering
\includegraphics[width=0.95\linewidth]{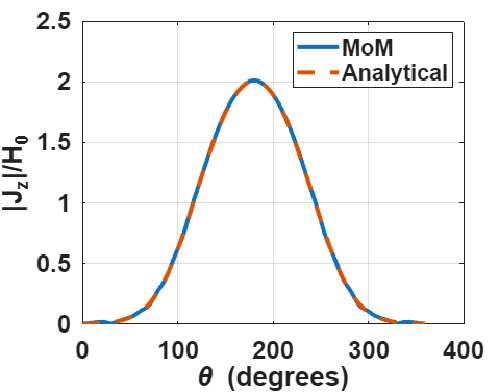}
\caption{Maginitude of the induced surface-current density for TM$_z$ scattering from a circular PEC cylinder with $R=2\lambda$.}
\label{fig:r2_tm_current1}
\end{figure}

To further validate the solver for a larger electrical size, the cylinder
radius was increased to $R=2\lambda$. Figure~\ref{fig:r2_tm_current1}
shows the surface-current comparison for TM$_z$ polarization. The MoM
current again matches the analytical result closely, indicating that the
EFIE formulation remains accurate for the larger cylinder.

\begin{figure}
\centerline{
\begin{tabular}{cc}
\includegraphics[width=0.49\linewidth]{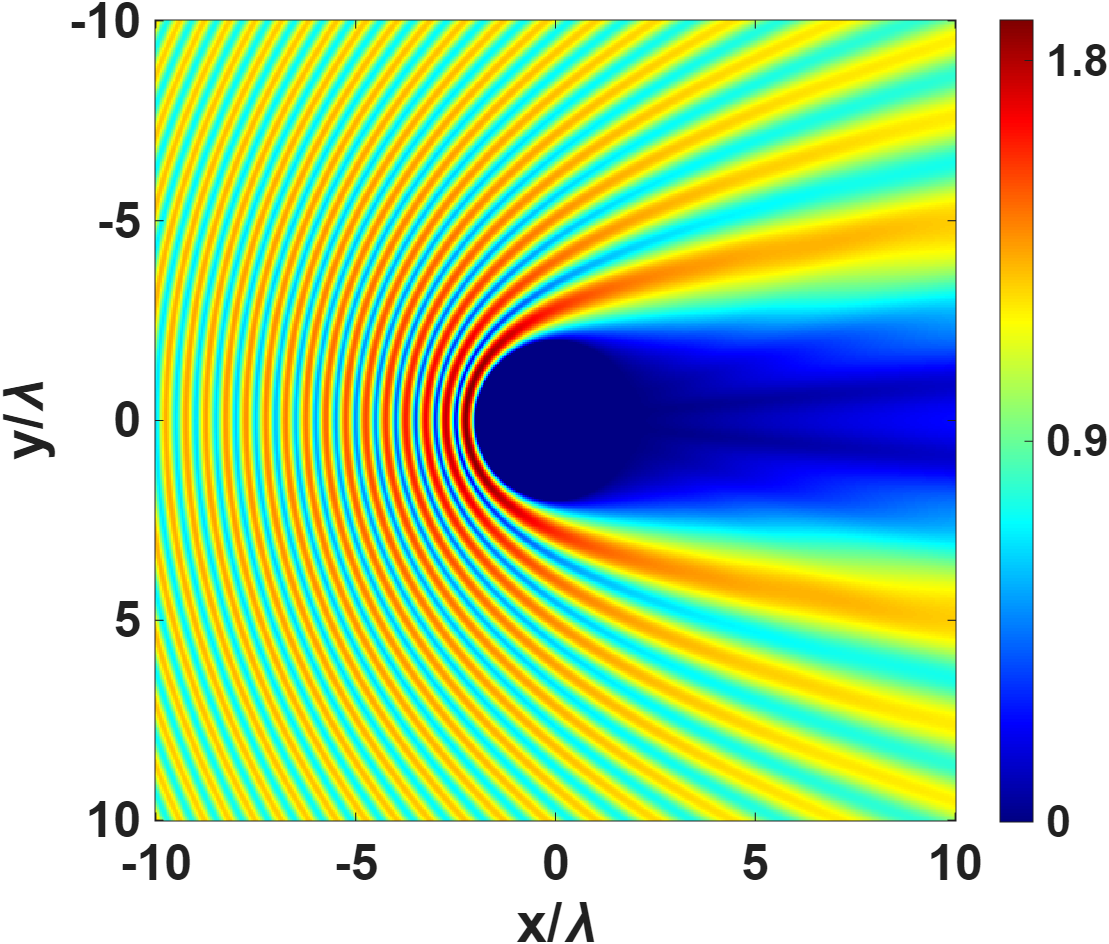} &
\includegraphics[width=0.49\linewidth]{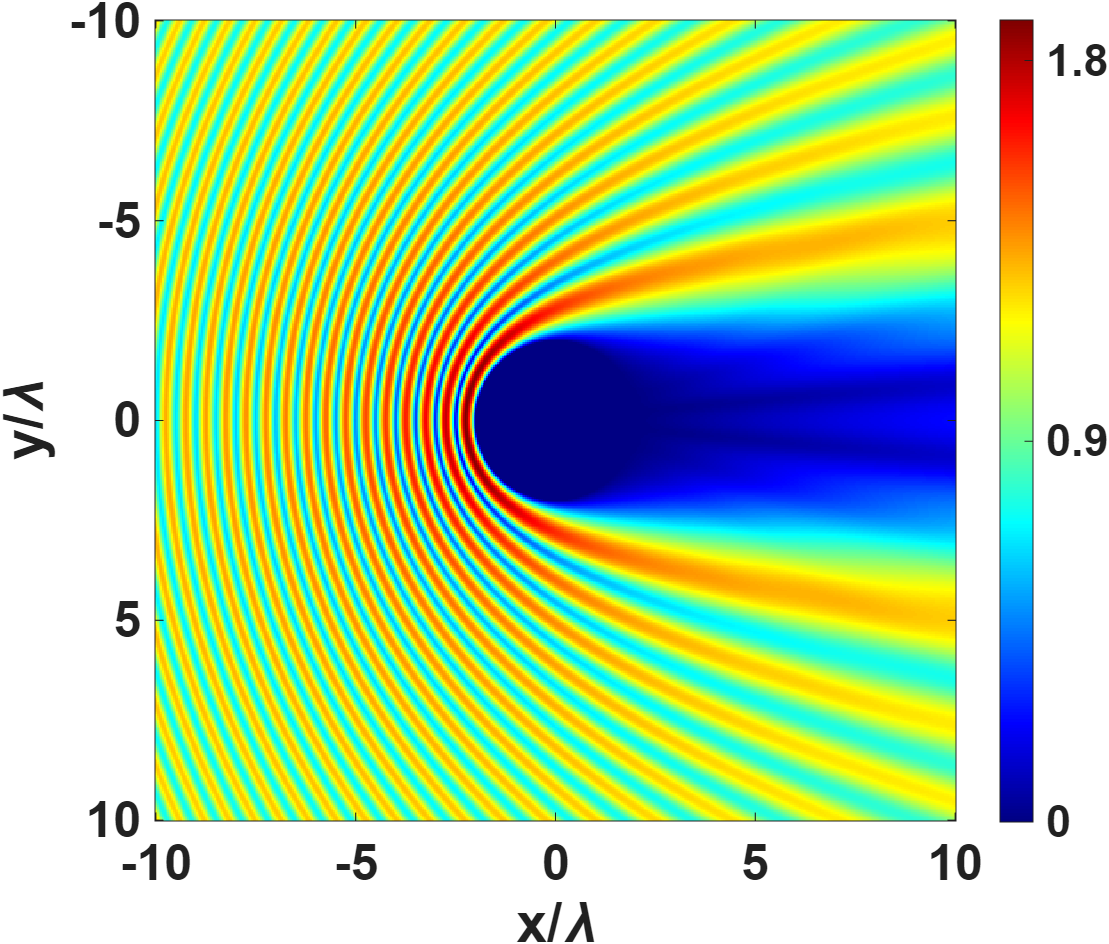} \\
{\centering\footnotesize (a) Analytical $|E_z^{\mathrm{tot}}|$} & {\centering\footnotesize (b) MoM $|E_z^{\mathrm{tot}}|$} \\
\end{tabular}
}
\vspace{2mm}
\centerline{
\begin{tabular}{cc}
\includegraphics[width=0.49\linewidth]{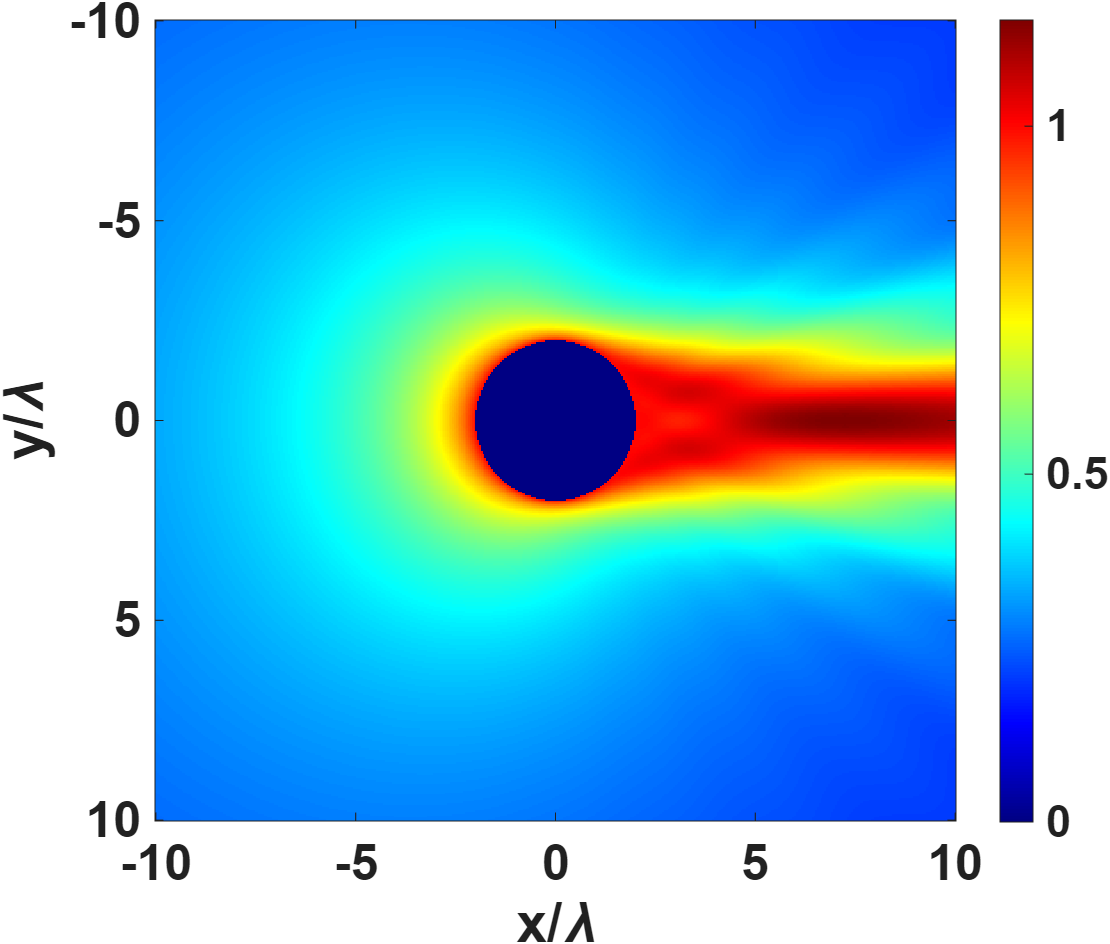} &
\includegraphics[width=0.49\linewidth]{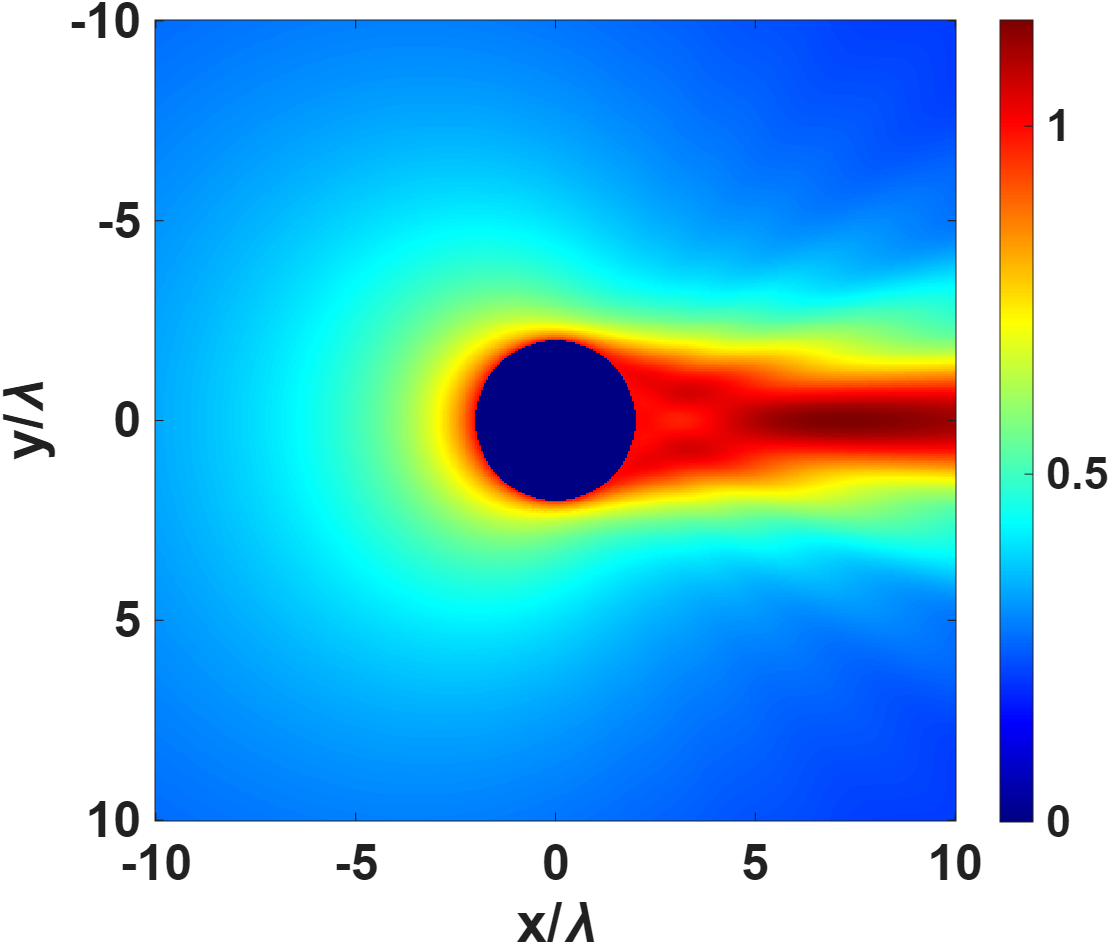} \\
{\centering\footnotesize (c) Analytical $|E_z^{\mathrm{scat}}|$} & {\centering\footnotesize (d) MoM $|E_z^{\mathrm{scat}}|$} \\
\end{tabular}
}
\caption{Analytical and MoM near-field comparison for TM$_z$ scattering
from a circular PEC cylinder with $R=2\lambda$.}
\label{fig:r2_tm_fields}
\vspace{-5mm}
\end{figure}

The field comparisons are shown in Fig.~\ref{fig:r2_tm_fields}. Relative
to the $R=\lambda$ case, the larger cylinder produces a wider shadow
region and more rapid field variation around the illuminated side of the
boundary. Nevertheless, the MoM total and scattered fields remain in
strong agreement with the analytical results. This confirms that the
TM$_z$ formulation retains its accuracy as the cylinder becomes electrically larger.

\subsubsection{TE$_z$ Polarization}

\begin{figure}[b]
\centering
\includegraphics[width=0.95\linewidth]{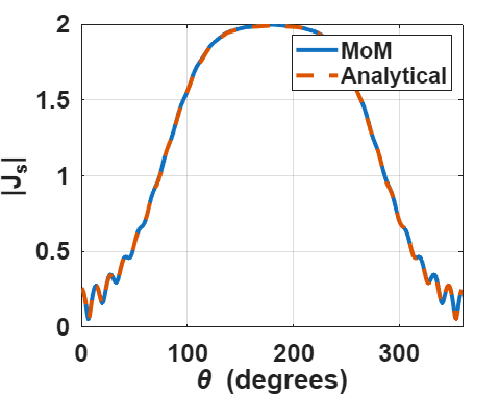}
\caption{Maginitude of the induced surface-current density for TE$_z$ scattering from a circular PEC cylinder with $R=2\lambda$.}
\label{fig:r2_te_current1}
\end{figure}

Finally, TE$_z$ scattering was studied for the circular PEC cylinder with
$R=2\lambda$. The surface-current comparison in
Fig.~\ref{fig:r2_te_current1} again shows close agreement between the
analytical and MoM results. As expected, the current distribution becomes
more oscillatory for the larger electrical size.

\begin{figure}[t]
\centerline{
\begin{tabular}{cc}
\includegraphics[width=0.49\linewidth]{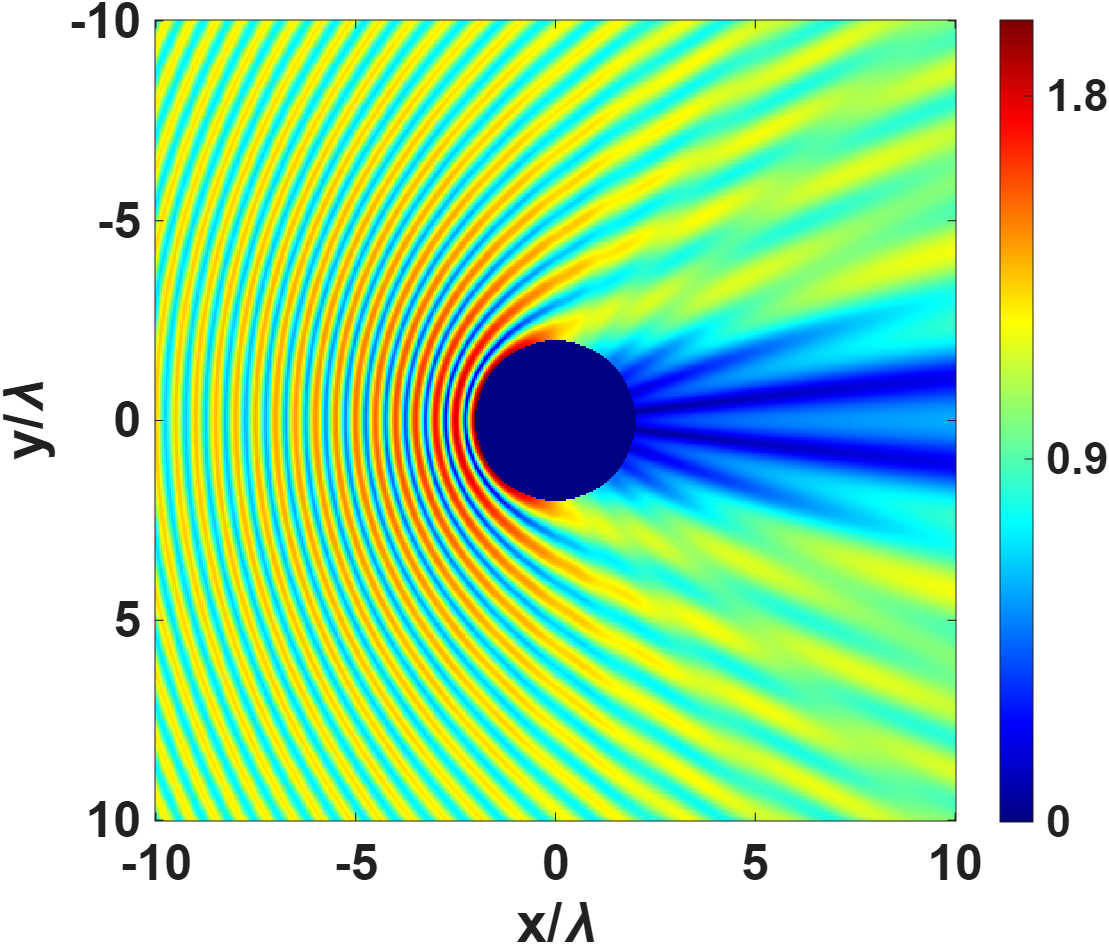} &
\includegraphics[width=0.49\linewidth]{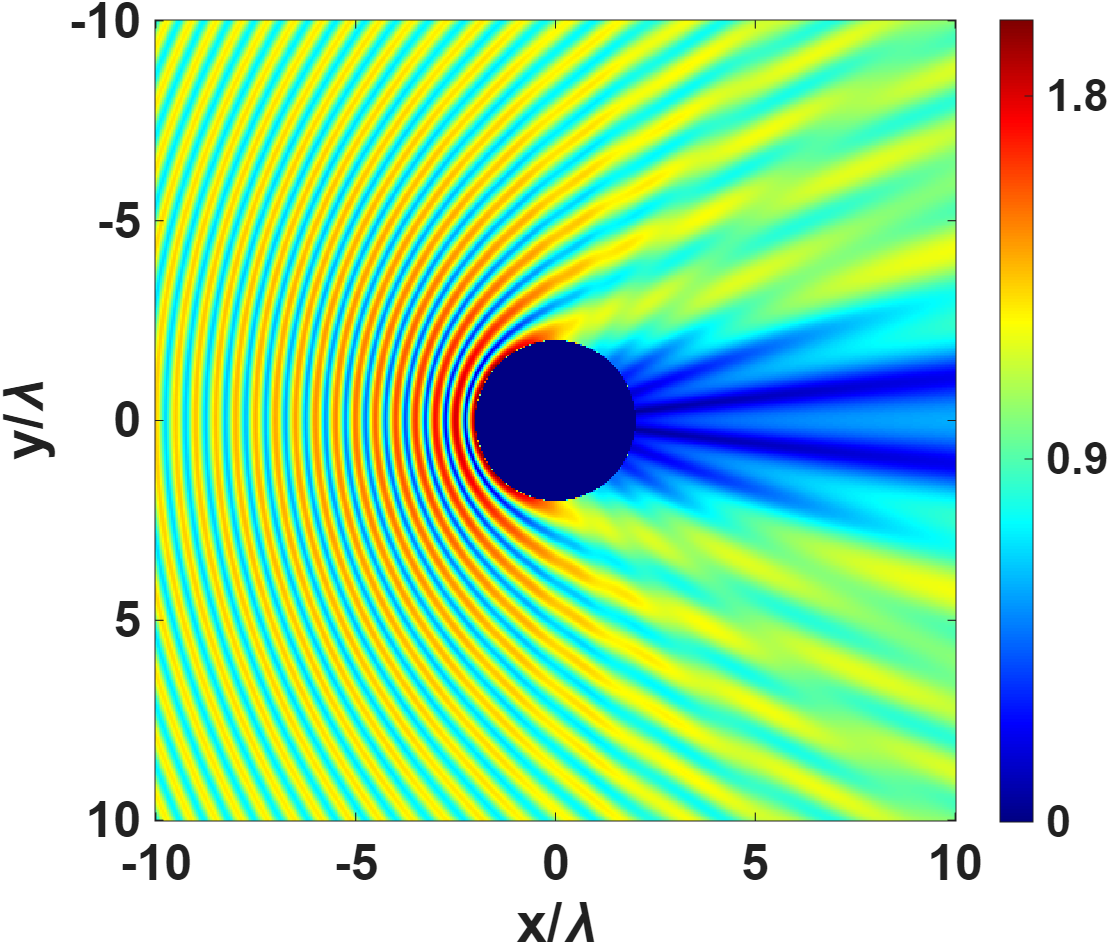} \\
{\centering\footnotesize (a) Analytical $|H_z^{\mathrm{tot}}|$} & {\centering\footnotesize (b) MoM $|H_z^{\mathrm{tot}}|$} \\
\end{tabular}
}
\vspace{2mm}
\centerline{
\begin{tabular}{cc}
\includegraphics[width=0.49\linewidth]{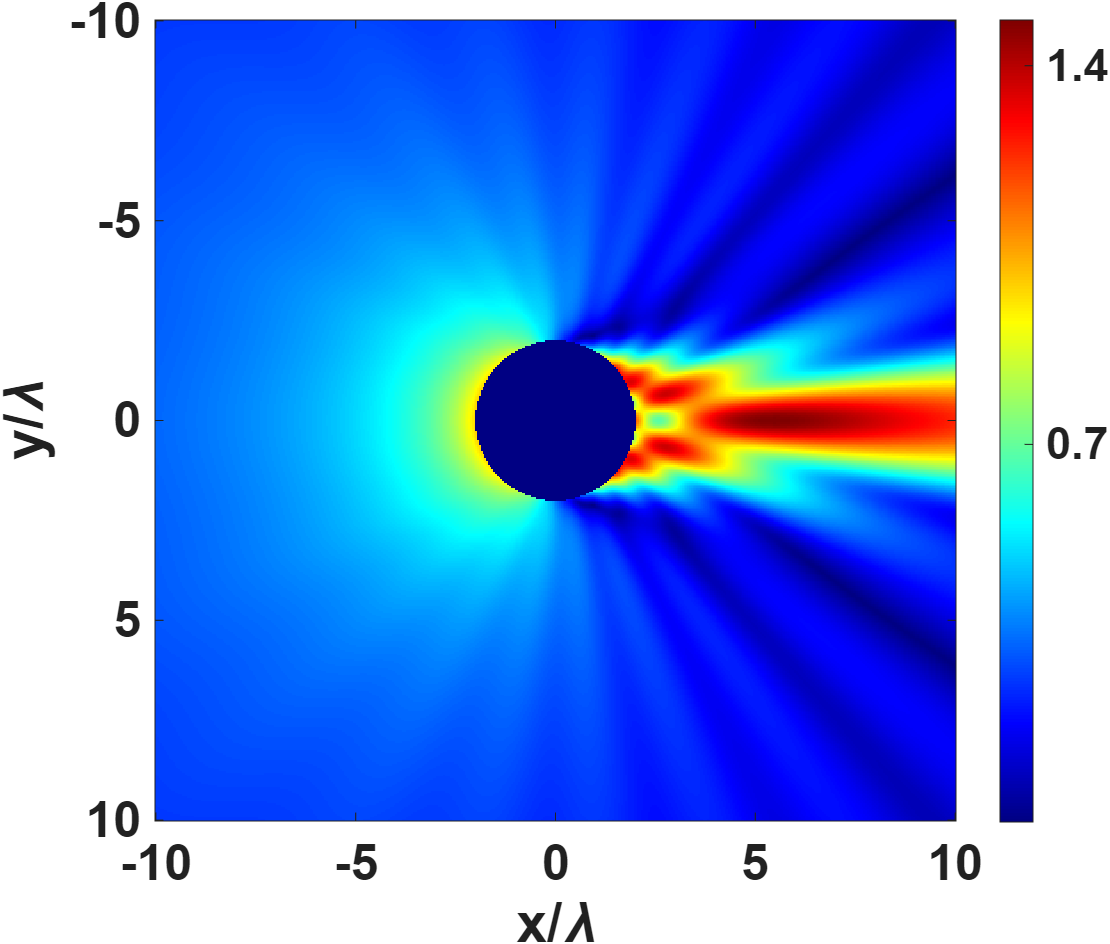} &
\includegraphics[width=0.49\linewidth]{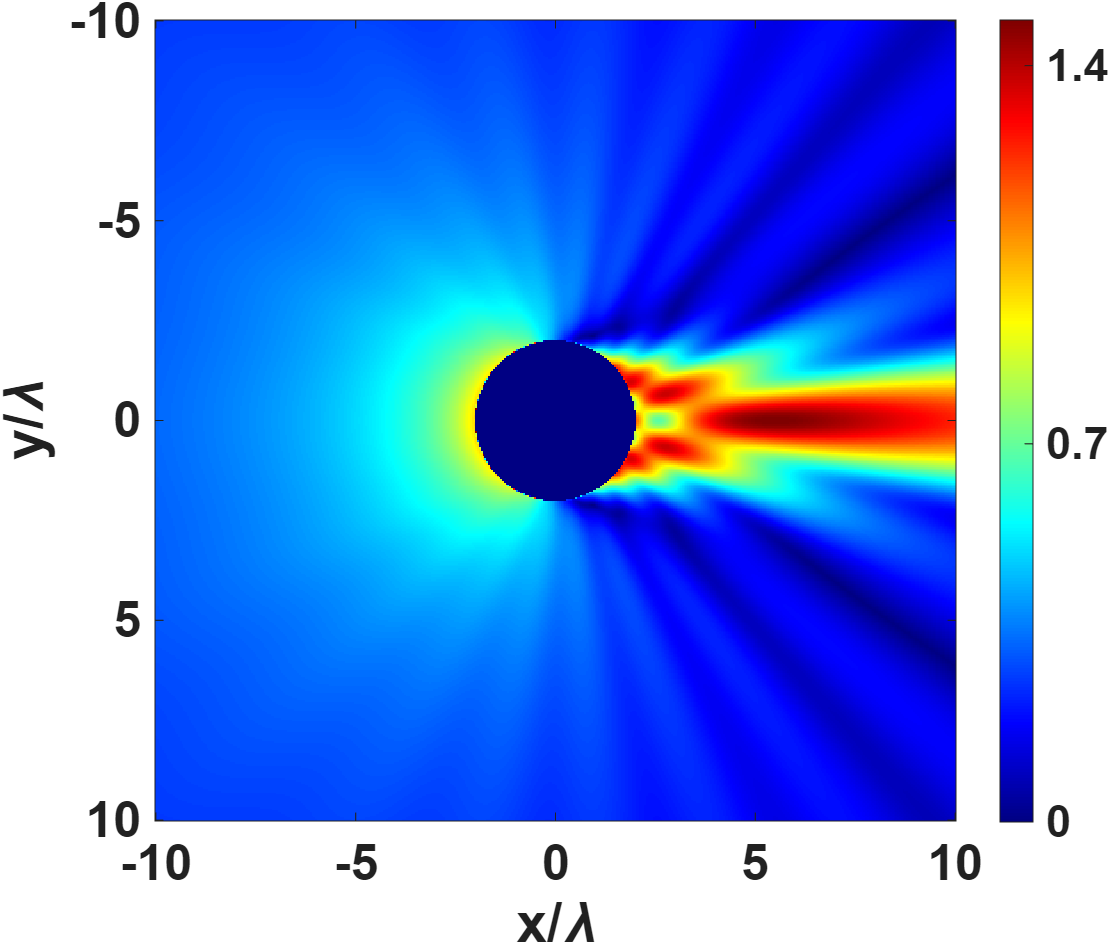} \\
{\centering\footnotesize (c) Analytical $|H_z^{\mathrm{scat}}|$} & {\centering\footnotesize (d) MoM $|H_z^{\mathrm{scat}}|$} \\
\end{tabular}
}
\caption{Analytical and MoM near-field comparison for TE$_z$ scattering
from a circular PEC cylinder with $R=2\lambda$.}
\label{fig:r2_te_fields}
\vspace{-5mm}
\end{figure}

Figure~\ref{fig:r2_te_fields} shows the total-field and scattered-field
comparisons. The larger cylinder produces stronger forward shadowing and
a more structured scattered-field pattern than in the $R=\lambda$ case.
Even so, the MoM and analytical results remain visually almost
indistinguishable. This final validation case confirms that the TE$_z$
MFIE implementation is also accurate for electrically larger PEC
cylinders.

\subsection{Square PEC Cylinder with Cross Section $3\lambda \times 3\lambda$}

After the circular-cylinder validation, the MoM solver was applied to a
$3\lambda \times 3\lambda$ square PEC cylinder as a non-circular test case.
No analytical reference solution is used here; instead, the results are
checked by examining the expected current concentration, shadowing, and
edge-scattering behavior.

\begin{figure}[t]
\centering
\begin{subfigure}[b]{0.9\linewidth}
    \centering
    \includegraphics[width=\linewidth]{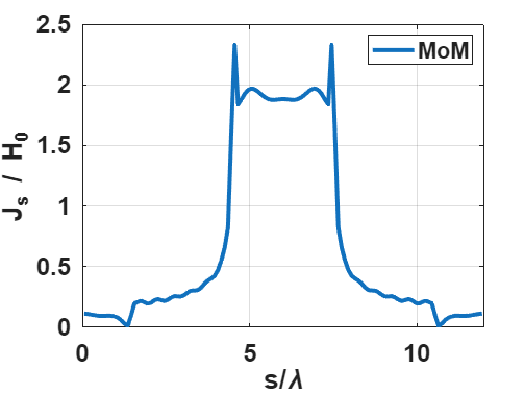}
    \caption{TM$_z$ polarization}
    \label{fig:square_tm_current}
\end{subfigure}
\vskip\baselineskip
\begin{subfigure}[b]{0.9\linewidth}
    \centering
    \includegraphics[width=\linewidth]{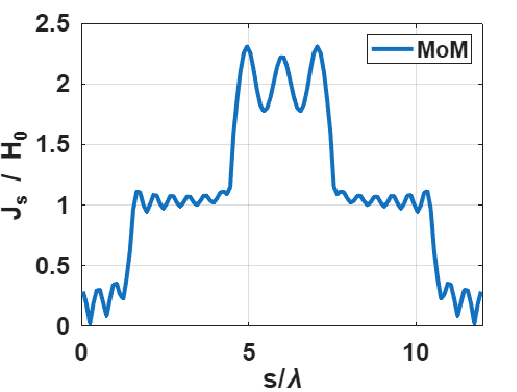}
    \caption{TE$_z$ polarization}
    \label{fig:square_te_current}
\end{subfigure}
\caption{Magnitude of induced surface-current for scattering from a $3\lambda\times3\lambda$ square PEC cylinder.}
\label{fig:square_current}
\end{figure}

\begin{figure}[h!]
\centerline{
\begin{tabular}{cc}
\includegraphics[width=0.49\linewidth]{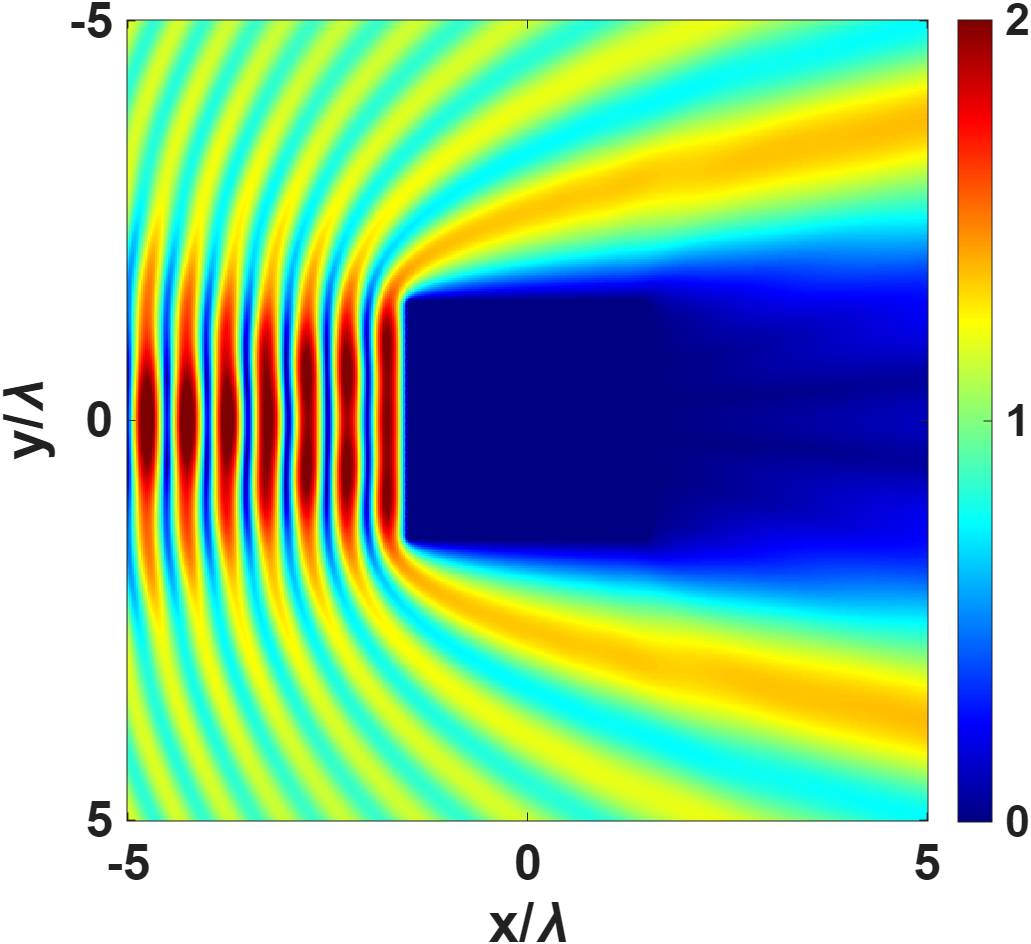} &
\includegraphics[width=0.49\linewidth]{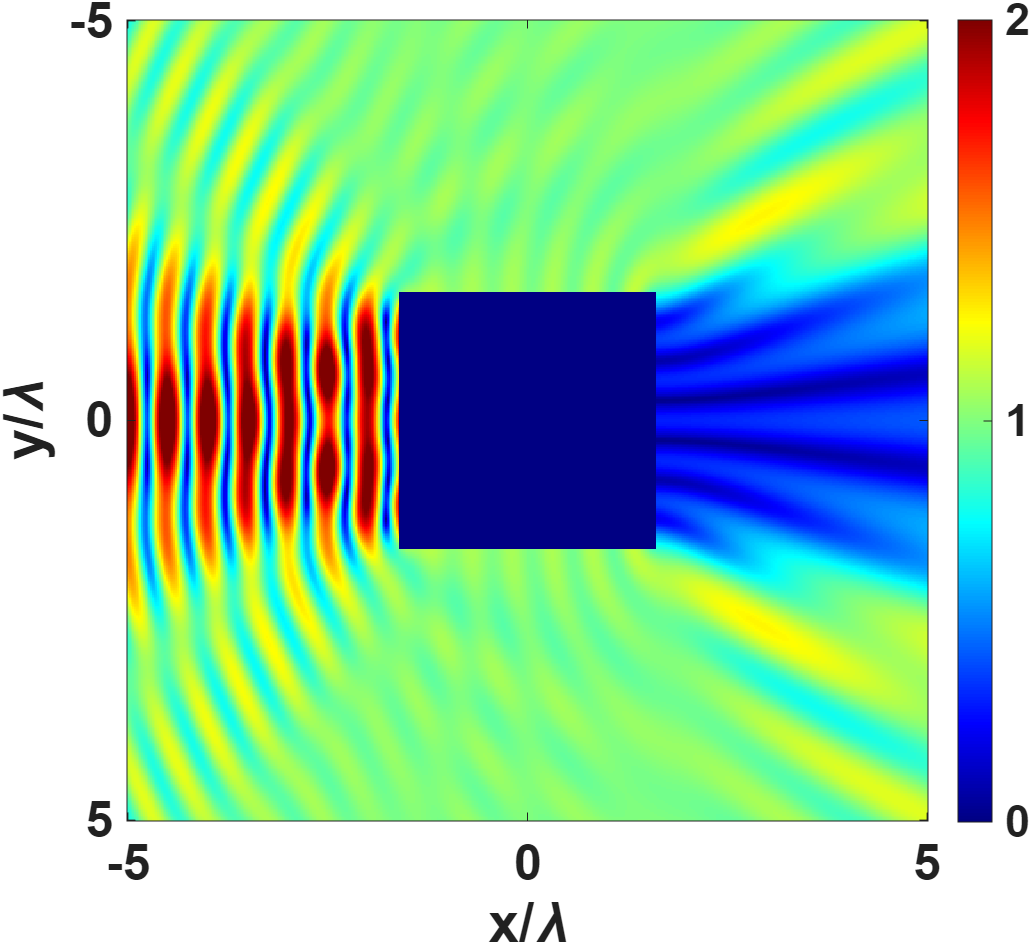} \\
{\centering\footnotesize (a) TM$_z$ $|E_z^{\mathrm{scat}}|$} &
{\centering\footnotesize (b) TE$_z$ $|H_z^{\mathrm{scat}}|$}
\end{tabular}
}
\vspace{2mm}
\centerline{
\begin{tabular}{cc}
\includegraphics[width=0.49\linewidth]{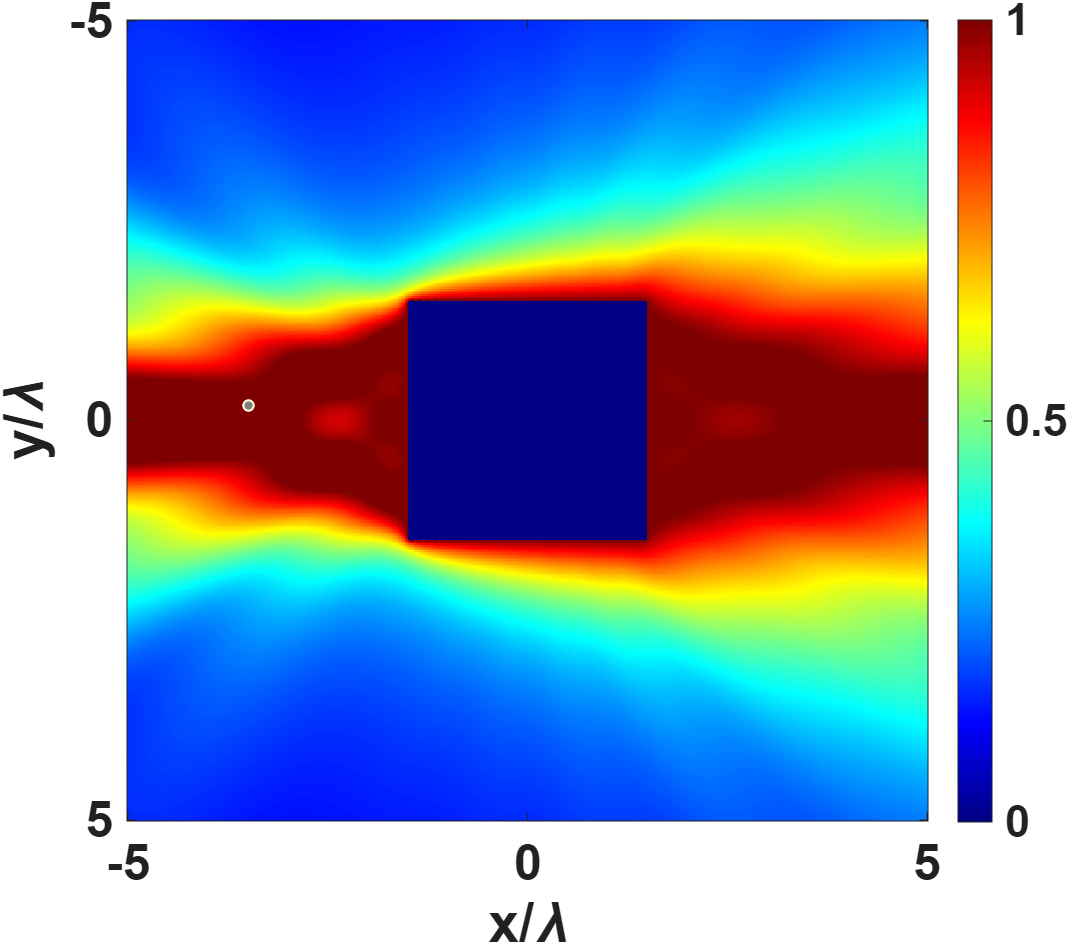} &
\includegraphics[width=0.49\linewidth]{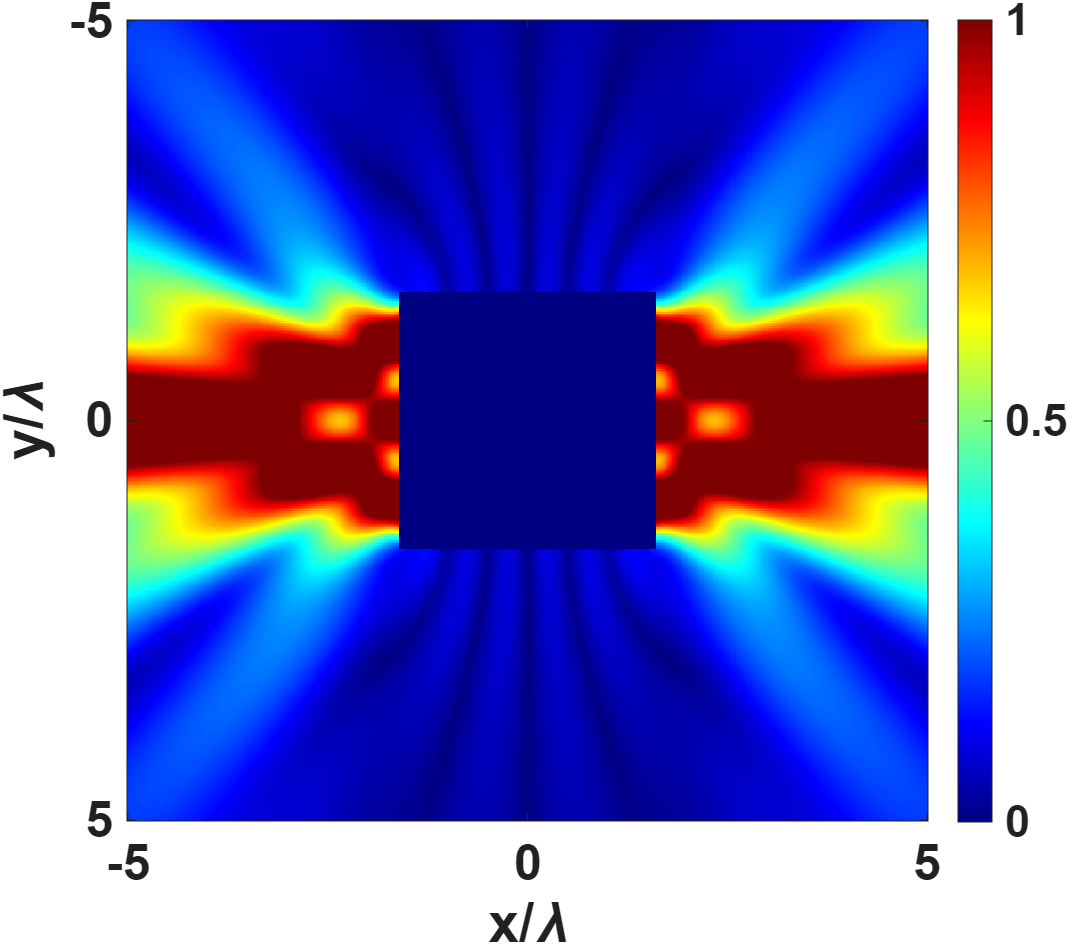} \\
{\centering\footnotesize (c) TM$_z$ $|E_z^{\mathrm{tot}}|$} &
{\centering\footnotesize (d) TE$_z$ $|H_z^{\mathrm{tot}}|$}
\end{tabular}
}
\caption{MoM near-field results for scattering from a
$3\lambda\times3\lambda$ square PEC cylinder.}
\label{fig:square_results}
\vspace{-5mm}
\end{figure}

Figure~\ref{fig:square_tm_current} shows the normalized TM$_z$ surface
current. The current is strongest on the illuminated face and near the
front corners, which is expected because the incident field directly
excites that side and the sharp edges cause rapid current variation.
Figure~\ref{fig:square_te_current} shows the TE$_z$ current, where a more
oscillatory contour variation is observed due to edge diffraction and the
normal-derivative behavior of the MFIE kernel.
Figure~\ref{fig:square_results}(a) shows the TM$_z$ scattered field, while
Fig.~\ref{fig:square_results}(b) shows the TE$_z$ scattered field. In both
plots, the scattered field is no longer smooth and symmetric as in the
circular-cylinder case. Instead, the flat faces and sharp corners of the
square cylinder produce stronger directional features and visible edge
diffraction. This behavior is especially noticeable in the TE$_z$ case,
where the scattered field contains more angular variation around the
object.

Figures~\ref{fig:square_results}(c) and
\ref{fig:square_results}(d) show the corresponding total fields for
TM$_z$ and TE$_z$ polarizations. In both cases, the incident wave is
strongly disturbed near the illuminated face of the square cylinder, and a
clear shadow region forms behind the PEC object. The total-field plots
therefore show that the MoM solution captures both the blockage effect of
the conductor and the geometry-dependent scattering caused by the square
cross section.

\section{Conclusion}

A two-dimensional method-of-moments solver was developed for TM$_z$ and
TE$_z$ scattering from infinitely long PEC cylinders. Starting from the
scalar Helmholtz equation and the two-dimensional outgoing Green's
function, the TM$_z$ problem was formulated using the EFIE for the
$z$-directed surface current, while the TE$_z$ problem was formulated
using the MFIE for the tangential surface current. The unknown currents were expanded using pulse basis functions, and the integral equations were
discretized by point matching at the segment centers.The implementation was validated using circular PEC cylinders with
radii $R=\lambda$ and $R=2\lambda$. For both polarizations, the computed
surface-current magnitudes, total near fields, and scattered near fields
showed close agreement with the analytical cylindrical-series solutions.
The small error levels in the validation cases indicate that the MoM code
correctly captures the main surface-current and near-field behavior for
canonical PEC-cylinder scattering problems.

The validated solver was then applied to a $3\lambda\times3\lambda$
square PEC cylinder. Since this geometry was not compared with an
analytical solution, the results were interpreted physically. The square
case showed stronger localized current variation near the illuminated
corners and more directional scattered-field patterns than the circular
case. These features are consistent with the presence of flat faces,
sharp corners, and edge diffraction, and they demonstrate that the same
MoM framework can be applied to non-circular conducting geometries.

Several improvements could be made in future work. A more systematic
mesh-refinement study would help quantify convergence with respect to
boundary segmentation density. Local refinement near corners could improve
the current representation for polygonal scatterers. Higher-order basis
functions and more accurate singular-integration schemes could further
increase the accuracy of the EFIE and MFIE discretizations. The work could
also be extended by computing far-field bistatic scattering widths and by
using fast algorithms, such as the fast multipole method, for larger
problems with many more boundary unknowns.



\end{document}